\documentclass[journal,twocolumn,10pt,twoside]{IEEEtran}
\normalsize

\ifCLASSINFOpdf
\else
\fi

\usepackage{amsmath}
\usepackage{amssymb}
\usepackage{cite}
\usepackage{graphicx}
\usepackage{algorithm}
\usepackage{algpseudocode}
\usepackage{subfigure}
\usepackage{multirow} 
\usepackage{longtable}
\usepackage{threeparttable}
\usepackage{caption3}
\usepackage{array}
\usepackage{cases}
\usepackage{color}
\usepackage[T1]{fontenc}
\usepackage[utf8]{inputenc}
\usepackage{authblk}
\usepackage[T1]{fontenc}
\usepackage{aecompl}

\newtheorem{remark}{Remark}

\newcommand\mycom[2]{\genfrac{}{}{0pt}{}{#1}{#2}}

\begin{document}
\title{Waveform Design for Wireless Power Transfer}
\author{Bruno Clerckx and Ekaterina Bayguzina
\thanks{The authors are with the EEE department at Imperial College London, London SW7 2AZ, United Kingdom (email: \{b.clerckx,ekaterina.bayguzina08\}@imperial.ac.uk). This work has been partially supported by the EPSRC of the UK under grants EP/K502856/1, EP/L504786/1 and EP/P003885/1. The material in this paper was presented in part at the IEEE ISWCS 2015 \cite{Clerckx:2015}.\\
Manuscript version \today}}
\maketitle

\begin{abstract} Far-field Wireless Power Transfer (WPT) has attracted significant attention in recent years. Despite the rapid progress, the emphasis of the research community in the last decade has remained largely concentrated on improving the design of energy harvester (so-called rectenna) and has left aside the effect of transmitter design. In this paper, we study the design of transmit waveform so as to enhance the DC power at the output of the rectenna. We derive a tractable model of the non-linearity of the rectenna and compare with a linear model conventionally used in the literature. We then use those models to design novel multisine waveforms that are adaptive to the channel state information (CSI). Interestingly, while the linear model favours narrowband transmission with all the power allocated to a single frequency, the non-linear model favours a power allocation over multiple frequencies. Through realistic simulations, waveforms designed based on the non-linear model are shown to provide significant gains (in terms of harvested DC power) over those designed based on the linear model and over non-adaptive waveforms. We also compute analytically the theoretical scaling laws of the harvested energy for various waveforms as a function of the number of sinewaves and transmit antennas. Those scaling laws highlight the benefits of CSI knowledge at the transmitter in WPT and of a WPT design based on a non-linear rectenna model over a linear model. Results also motivate the study of a promising architecture relying on large-scale multisine multi-antenna waveforms for WPT. As a final note, results stress the importance of modeling and accounting for the non-linearity of the rectenna in any system design involving wireless power.  
\end{abstract}


\IEEEpeerreviewmaketitle

\section{Introduction}


\IEEEPARstart{W}{ireless} Power Transfer (WPT) via radio-frequency radiation has a long history that is nowadays attracting more and more attention. RF radiation has indeed become a viable source for energy harvesting with clear applications in Wireless Sensor Networks (WSN) and Internet of Things (IoT) \cite{Visser:2013}. 
The major challenge facing far-field wireless power designers is to find ways to increase the DC power level at the output of the rectenna without increasing the transmit power, and for devices located tens to hundreds of meters away from the transmitter. To that end, the vast majority of the technical efforts in the literature have been devoted to the design of efficient rectennas, a.o. \cite{Visser:2013,Pinuela:2013,Hagerty:2004}. A rectenna harvests ambient electromagnetic energy, then rectifies and filters it (using a diode and a low pass filter). The recovered DC power then either powers a low power device directly, or is stored in a super capacitor for higher power low duty-cycle operation.
\par Interestingly, the overall RF-to-DC conversion efficiency of the rectenna is not only a function of its design but also of its input waveform. However, the waveform design has received less attention \cite{Trotter:2009,Boaventura:2011,Collado:2014}. In \cite{Trotter:2009,Boaventura:2011}, a multisine signal excitation is shown through analysis, simulations and measurements to enhance the DC power and RF-to-DC conversion efficiency over a single sinewave signal. In \cite{Collado:2014}, various input waveforms (OFDM, white noise, chaotic) are considered and experiments show that waveforms with high peak to average power ratio (PAPR) increase RF-to-DC conversion efficiency. Even though those papers provide some useful insights into the impact of waveform design onto WPT performance, there are many limitations in the WPT waveform design literature: 1) there has not been any formal tool to design and optimize waveforms for WPT so far, 2) multipath fading (well known in wireless communications) has been ignored despite its tremendous impact on the received waveform at the input of the rectenna, 3) the Channel State Information (CSI) is assumed unknown to the transmitter, 4) the transmitter is commonly equipped with a single antenna and 5) a single rectenna is considered. 

\par In this paper we address the important problem of waveform design for WPT and tackle all the aforementioned limitations. We focus on multisine waveforms due to their tractability and usefulness in wireless communication systems. The contributions of the paper are summarized as follows.
\par First, we introduce a simple and tractable analytical model of the rectenna non-linearity through the second and higher order terms in the Taylor expansion of the diode characteristics. Comparison is made with a linear model, first introduced in \cite{Wetenkamp:1983} and nowadays popular in Simultaneous Wireless Information and Power Transfer (SWIPT), e.g. \cite{Zhou:2013} and subsequent works, that only accounts for the second order term.
\par Second, assuming perfect CSI at the Transmitter (CSIT) can be attained and making use of the rectenna model, we design multi-antenna multisine WPT waveform for transmission over a multipath channel. We formulate an optimization problem to adaptively change the waveform weights as a function of the CSI so as to maximize the rectenna output DC current. The global optimal phases of the multisine waveform weights are obtained in closed form while the amplitudes (not guaranteed to be global optimal) result from a non-convex posynomial maximization problem subject to a power constraint.
\par Third, the use of a linear or non-linear model of the rectenna is shown to lead to very different WPT system design. While the linear model favours a narrowband power allocation (over a single frequency), the non-linear model favours a wideband power allocation (over multiple frequencies). 
\par Fourth, the waveform design is generalized to multi-rectenna WPT and to account for PAPR constraints. The design results from a signomial maximization problem.
\par Fifth, scaling laws of the harvested energy with various waveforms are analytically derived as a function of the number of sinewaves $N$, the number of transmit antennas $M$ and the progagation conditions. We show for instance that in frequency-flat and frequency-selective channels and for a fixed transmit power constraint, the DC current at the output of the rectifier theoretically increases linearly with $N$ if the non-linear model is used for waveform design. Interestingly, while such a scaling law is achievable in frequency-flat channels without CSIT, it is achievable in frequency-selective channels only in the presence of CSIT. On the other hand, with a design based on the linear model, the DC current increases at most logarithmically with $N$. The results also motivate the usefulness of transmitting multisine waveforms and acquiring CSIT in WPT, especially in frequency-selective channels.  
\par Sixth, the waveforms designed for WPT, adaptive to the CSI and accounting for the rectifier non-linearity, are shown through realistic circuit evaluations to provide significant gains over state-of-the-art waveforms and over those optimized based on the linear model of the rectifier. Moreover, while the non-linear model is validated by circuit simulations, the linear model is shown to be inaccurate and unable to predict correctly the multisine waveform performance. 
\par As a main takeaway observation, the results highlight the importance of modeling and accounting for the non-linearity of the rectenna in any design and evaluations of system involving wireless power. 
\par \textit{Organization:} Section \ref{system_model} introduces the system model and section \ref{section_EH} models the rectenna. Section \ref{section_waveform} tackles the waveform optimization for a single and multiple rectennas, with and without PAPR constraints. Section \ref{section_scaling_law} analytically derives the scaling laws of the harvested energy. Section \ref{simulations} evaluates the performance and section \ref{conclusions} concludes the work.
\par \textit{Notations:} Bold lower and upper case letters stand for vectors and matrices respectively. A symbol not in bold font represents a scalar. $\left\|.\right\|$ and $\left\|.\right\|_F$ refer to the norm and Frobenius norm of a vector and matrix, respectively. $\mathcal{E}\left\{.\right\}$ is the expectation/averaging operator. $.^*$, $.^T$ and $.^H$ refer to the conjugate, transpose and conjugate transpose of a matrix, respectively. $\mathbf{1}_{N}$ and $\mathbf{0}_{N}$ refer to the $N\times 1$ vector with entries equal to $1$ and $0$, respectively. $\lambda_{max}$ refers to the largest eigenvalue of a matrix. $\log$ is in base $e$. $\left\vert{S}\right\vert$ is the cardinality of set $S$. $\stackrel{N\nearrow}{\approx}$ means approximately equal as $N$ grows large.

\section{WPT System Model}\label{system_model}
Consider a transmitter with $M$ antennas and $N$ sinewaves whose transmit signal at time $t$ on antenna $m$ is given by
\begin{align}\label{WPT_waveform}
x_m(t)=\Re\left\{\sum_{n=0}^{N-1}w_{n,m}e^{jw_n t}\right\}
\end{align}
with $w_{n,m}=s_{n,m}e^{j\phi_{n,m}}$ where $s_{n,m}$ and $\phi_{n,m}$ refer to the amplitude and phase of the $n^{th}$ sinewave at frequency $w_n$ on transmit antenna $m$, respectively. We assume for simplicity that the frequencies are evenly spaced, i.e.\ $w_n=w_0+n\Delta_w$ with $\Delta_w=2\pi\Delta_f$ the frequency spacing. The magnitudes and phases of the sinewaves can be collected into matrices $\mathbf{S}$ and $\mathbf{\Phi}$. The $(n,m)$ entry of $\mathbf{S}$ and $\mathbf{\Phi}$ write as $s_{n,m}$ and $\phi_{n,m}$, respectively. The $m^{th}$ column of $\mathbf{S}$ is denoted as $\mathbf{s}_m$. The transmitter is subject to a transmit power constraint $\sum_{m=1}^{M}\mathcal{E}\big\{\left|x_m\right|^2\big\}=\frac{1}{2}\left\|\mathbf{S}\right\|_F^2\leq P$.
Stacking up all transmit signals, we can write the transmit signal vector as $\mathbf{x}(t)=\Re\big\{\sum_{n=0}^{N-1}\mathbf{w}_{n}e^{jw_n t}\big\}$
where $\mathbf{w}_{n}=\big[\begin{array}{ccc}w_{n,1}&\ldots & w_{n,M}\end{array}\big]^T$\footnote{Note that $\mathbf{w}_{n}$ and $w_{n,m}$ should not be confused with $w_n$.}.

\par The multi-antenna transmitted sinewaves propagate through a multipath channel, characterized by $L$ paths whose delay, amplitude, phase and direction of departure (chosen with respect to the array axis) are respectively denoted as $\tau_l$, $\alpha_l$, $\xi_l$ and $\theta_l$, $l=1,\ldots,L$. We assume transmit antennas are closely located so that $\tau_l$, $\alpha_l$ and $\xi_l$ are the same for all transmit antennas (assumption of a narrowband balanced array) \cite{Clerckx:2013}. Denoting $\zeta_{n,m,l}=\xi_l+\Delta_{n,m,l}$ with $\Delta_{n,m,l}$ the phase shift between the $m^{th}$ transmit antenna and the first one\footnote{For a Uniform Linear Array (ULA), $\Delta_{n,m,l}=2\pi (m-1)\frac{d}{\lambda_n}\cos(\theta_l)$ where $d$ is the inter-element spacing, $\lambda_n$ the wavelength of the $n^{th}$ sinewave.}, the signal transmitted by antenna $m$ and received at the single-antenna receiver after multipath propagation can be written as
\begin{align}
y_m(t)&=\sum_{n=0}^{N-1} \sum_{l=0}^{L-1}s_{n,m}\alpha_l \cos(w_n(t-\tau_l)+\zeta_{n,m,l}+\phi_{n,m})\nonumber\\
&=\sum_{n=0}^{N-1}s_{n,m}A_{n,m} \cos(w_n t+\psi_{n,m})\label{received_signal_ant_m}
\end{align}
where the amplitude $A_{n,m}$ and the phase $\psi_{n,m}$ are such that
\begin{align}\label{A_nm_psi}
A_{n,m}e^{j \psi_{n,m}}&=A_{n,m}e^{j \left(\phi_{n,m}+\bar{\psi}_{n,m}\right)}=e^{j \phi_{n,m}}h_{n,m}
\end{align}
with $h_{n,m}=A_{n,m}e^{j \bar{\psi}_{n,m}}=\sum_{l=0}^{L-1}\alpha_l e^{j(-w_n\tau_l+\zeta_{n,m,l})}$ the frequency response of the channel of antenna $m$ at $w_n$. The vector channel is defined as $\mathbf{h}_n=\big[\begin{array}{ccc}h_{n,1}&\ldots & h_{n,M}\end{array}\big]$.
 
\par The total received signal comprises the sum of \eqref{received_signal_ant_m} over all transmit antennas, namely 
\begin{align}
y(t)=\sum_{n=0}^{N-1}X_n \cos(w_n t+\delta_{n})=\Re\left\{\sum_{n=0}^{N-1}\mathbf{h}_n\mathbf{w}_n e^{j w_n t}\right\}\label{y_t}
\end{align}
where $X_n e^{j \delta_{n}}=\sum_{m=1}^M s_{n,m}A_{n,m}e^{j \psi_{n,m}}=\mathbf{h}_n\mathbf{w}_n$.

\section{Analytical Model of the Rectenna}\label{section_EH}
We derive a simple and tractable model of the rectenna circuit and express the output DC current as a function of the waveform parameters. The model relies on several assumptions made to make the model tractable and be able to optimize the waveforms. Performance evaluations will be conducted in Section \ref{simulations} using a more accurate circuit simulator.

\subsection{Antenna Equivalent Circuit}\label{antenna_eq_circuit}
\begin{figure}
 \begin{minipage}[c]{.5\linewidth}
   \centerline{\includegraphics[width=0.9\columnwidth]{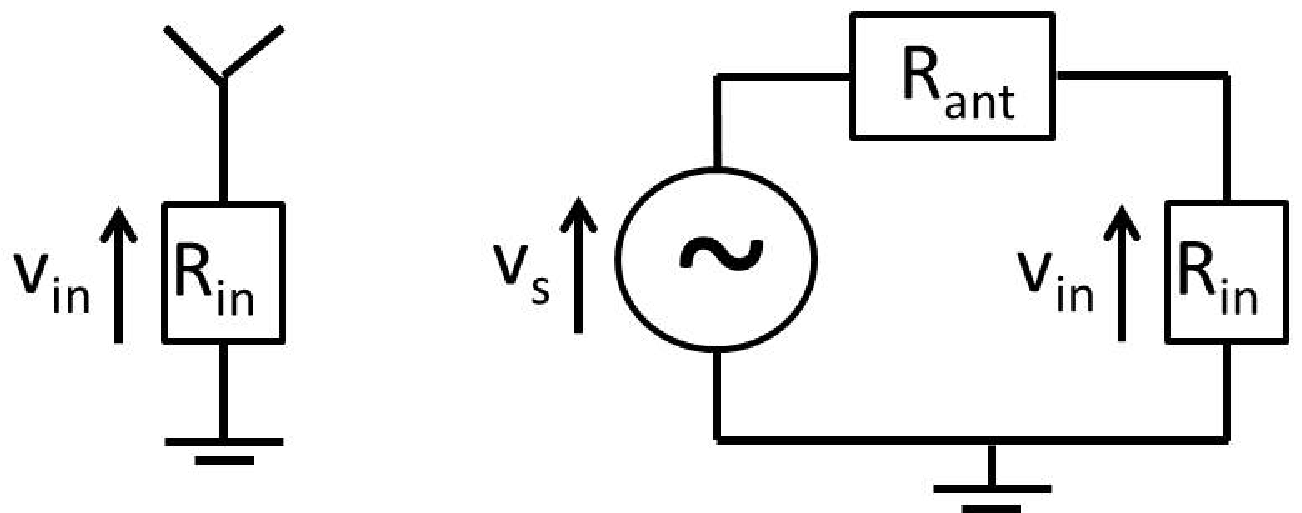}}
  \end{minipage}\hfill
 \begin{minipage}[c]{.5\linewidth}
   \centerline{\includegraphics[width=0.9\columnwidth]{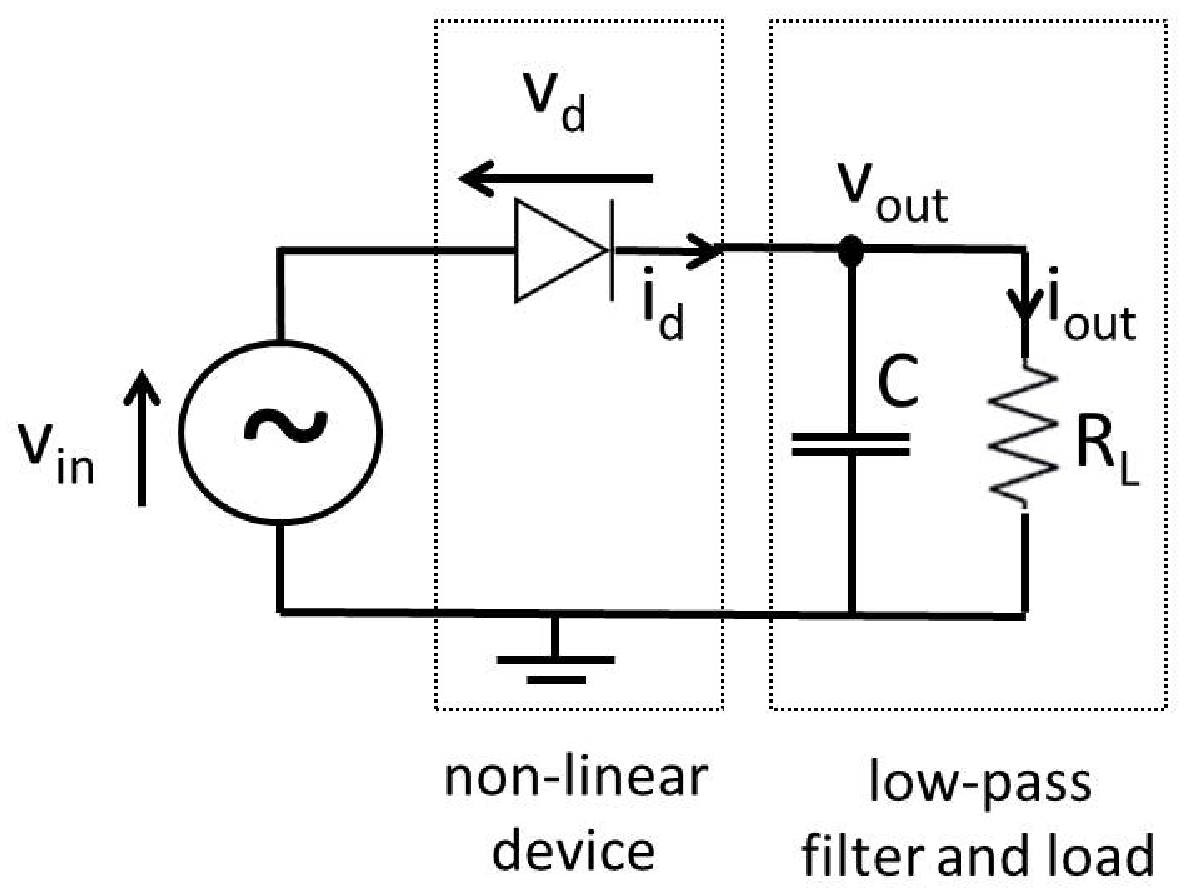}}
  \end{minipage}
  \caption{Antenna equivalent circuit (left) and a single diode rectifier (right).}
  \label{antenna_model}
\end{figure}

Assume a rectenna whose input impedance $R_{in}$ is connected to a receiving antenna as in Fig \ref{antenna_model}. The signal $y(t)$ impinging on the antenna has an average power $P_{av}=\mathcal{E}\big\{\left|y(t)\right|^2\big\}$. Following \cite{Curty:2005}, the antenna is assumed lossless and modeled as an equivalent voltage source $v_s(t)$ in series with an impedance $R_{ant}=50\Omega$, as illustrated in Fig \ref{antenna_model}. 
\par With perfect matching ($R_{in}=R_{ant}$), the received power $P_{av}$ is completely transferred to the rectenna's input impedance such that $P_{av}=\mathcal{E}\big\{\left|v_{in}(t)\right|^2\big\}/R_{in}$ where $v_{in}(t)$ is the rectifier's input voltage. Under perfect matching, $v_{in}(t)$ is half of $v_s(t)$ and both can be related to the received signal $y(t)$ as $v_s(t)=2 y(t) \sqrt{R_{ant}}$ and $v_{in}(t)=y(t) \sqrt{R_{ant}}$, such that $P_{av}=\mathcal{E}\big\{\left|v_{in}(t)\right|^2\big\}/R_{in}=\mathcal{E}\big\{\left|y(t)\right|^2\big\}R_{ant}/R_{in}=\mathcal{E}\big\{\left|y(t)\right|^2\big\}$.
We also assume that the antenna noise is too small to be harvested so as no antenna noise term is added and $v_{in}(t)$ is delivered as such to the rectifier. 


\subsection{Rectifier and Diode Non-Linearity}\label{rectifier_subsection}
A rectifier is always made of a non-linear device (e.g.\ diode) followed by a low pass filter (LPF) with load \cite{Pinuela:2013,Trotter:2009,Boaventura:2011}. A simplified rectifier circuit is illustrated in Fig \ref{antenna_model}. We assume that its input impedance has been perfectly matched to the antenna impedance. 


The current $i_d(t)$ flowing through an ideal diode (neglecting its series resistance) relates to the voltage drop across the diode $v_d(t)=v_{in}(t)-v_{out}(t)$ as $i_d(t)=i_s\big(e^{\frac{v_d(t)}{n v_t}}-1\big)$
where $i_s$ is the reverse bias saturation current, $v_t$ is the thermal voltage, $n$ is the ideality factor (assumed equal to $1.05$). In order to express the non-linearity of the diode, we take a Taylor expansion of the exponential function around a fixed operating voltage drop $v_d=a$ such that the diode current can be equivalently written as
\begin{equation}\label{taylor}
i_d(t)=\sum_{i=0}^\infty k_i (v_d(t)-a)^i=\sum_{i=0}^\infty k_i (v_{in}(t)-v_{out}(t)-a)^i,
\end{equation}
where $k_0=i_s\big(e^{\frac{a}{n v_t}}-1\big)$ and $k_i=i_s\frac{e^{\frac{a}{n v_t}}}{i!\left(n v_t\right)^i}$, $i=1,\ldots,\infty$. The Taylor series expansion model is a small signal model that is valid only for the non-linear operation region of the diode. If the input voltage amplitude becomes large, the diode will be driven into the large signal operation region where the diode behaviour is dominated by the diode series resistance and the I-V relationship is linear \cite{OptBehaviour}.
 
\par As such, it is not easy to infer from \eqref{taylor} the exact dependencies of the diode current on the waveform parameters since both $v_{in}(t)$ and $v_{out}(t)$ will depend and fluctuate over time as a function of the waveform. Nevertheless, assuming a steady-state response, an ideal rectifier would deliver a constant (over time) output voltage $v_{out}$ whose level would depend on the peaks of the input voltage $v_{in}(t)$ \cite{Curty:2005}. As a consequence, the output current delivered to the load $i_{out}$ would also be constant. 
In this ideal rectifier, since $v_{out}$ is a constant (we drop the time dependency), a suitable choice of the operating voltage drop $a$ would be $a=\mathcal{E}\left\{v_{in}(t)-v_{out}\right\}=-v_{out}$ since $\mathcal{E}\left\{v_{in}(t)\right\}=\sqrt{R_{ant}}\mathcal{E}\left\{y(t)\right\}=0$. Under such assumptions, \eqref{taylor} can simply be written as
\begin{equation}\label{taylor_bis}
i_d(t)=\sum_{i=0}^\infty k_i v_{in}(t)^i=\sum_{i=0}^\infty k_i R_{ant}^{i/2} y(t)^i,
\end{equation}
which makes the dependency between the diode current $i_d(t)$, the received waveform $y(t)$ and therefore the transmitted waveforms $\left\{x_m(t)\right\}$ much more explicit. 
\par The problem at hand will be the design of $\left\{x_m(t)\right\}$ such that the output DC current is maximized. Under the ideal rectifier assumption, the current delivered to the load in a steady-state response is constant and given by $i_{out}=\mathcal{E}\left\{i_d(t)\right\}$, i.e. the average over time of the current flowing through the diode. In order to make the optimization tractable, we truncate the Taylor expansion to the $n_o^{th}$ order. We consider two models: a non-linear model that truncates the Taylor expansion to the $n_o^{th}$ order but retains the fundamental non-linear behaviour of the diode and a linear model that truncates to the second order term and ignores the non-linearity.

\subsection{A Non-Linear Model}
After truncation, the output DC current approximates as 
\begin{equation}\label{diode_model}
i_{out}=\mathcal{E}\left\{i_d(t)\right\}\approx \sum_{i=0}^{n_o} k_i R_{ant}^{i/2} \mathcal{E}\left\{y(t)^i\right\}.
\end{equation}
\begin{table*}
\begin{align}\label{yt_2}
y(t)^2&=\frac{1}{2}\sum_{n_0,n_1}X_{n_0}X_{n_1}\left[\cos(w^{++}t+\delta^{++})+\cos(w^{+-}t+\delta^{+-})\right],\\
y(t)^3&=\frac{1}{4}\sum_{n_0,n_1,n_2}X_{n_0}X_{n_1}X_{n_2}\left[\cos(w^{+++}t+\delta^{+++})+\cos(w^{++-}t+\delta^{++-})+\cos(w^{+-+}t+\delta^{+-+})+\cos(w^{+--}t+\delta^{+--})\right],\label{yt_3}\\
y(t)^4&=\frac{1}{8}\sum_{\mycom{n_0,n_1,}{n_2,n_3}}X_{n_0}X_{n_1}X_{n_2}X_{n_3}\left[\cos(w^{++++}t+\delta^{++++})+\cos(w^{++--}t+\delta^{++--})+\cos(w^{+++-}t+\delta^{+++-})+\cos(w^{++-+}t+\delta^{++-+})\right.\nonumber\\
&\hspace{2cm}\left.+\cos(w^{+-++}t+\delta^{+-++})+\cos(w^{+---}t+\delta^{+---})+\cos(w^{+-+-}t+\delta^{+-+-})+\cos(w^{+--+}t+\delta^{+--+})\right].\label{yt_4}
\end{align}
\hrulefill
\end{table*}
Applying \eqref{y_t} to \eqref{diode_model} involves the computation of $y(t)^i$, illustrated in \eqref{yt_2}, \eqref{yt_3} and \eqref{yt_4} for $i=2,3,4$. In order to simplify the notations, \eqref{yt_2} makes use of $w^{++}$ and $\delta^{++}$ to denote $w_{n_0}+w_{n_1}$ and $\delta_{n_0}+\delta_{n_1}$, respectively. Hence the sign of $\left\{w_{n_0},w_{n_1}\right\}$ and $\left\{\delta_{n_0},\delta_{n_1}\right\}$ is reflected as a superscript. Similarly, $w^{+-}=w_{n_0}-w_{n_1}$ and $\delta^{+-}=\delta_{n_0}-\delta_{n_1}$. In \eqref{yt_3} and \eqref{yt_4}, we use the same convention, e.g.\ $w^{++++}=w_{n_0}+w_{n_1}+w_{n_2}+w_{n_3}$, $w^{++--}=w_{n_0}+w_{n_1}-w_{n_2}-w_{n_3}$, etc. Averaging over time, we get an approximation of the DC component of the current at the output of the rectifier (and the low-pass filter) with a multisine excitation over a multipath channel as 
\begin{equation}\label{diode_model_2}
i_{out}\approx k_0+\sum_{i \hspace{0.1cm}\textnormal{even}, i\geq 2}^{n_o} k_i R_{ant}^{i/2} \mathcal{E}\left\{y(t)^i\right\} 
\end{equation}
with $\mathcal{E}\left\{y(t)^2\right\}$, $\mathcal{E}\left\{y(t)^4\right\}$ and $\mathcal{E}\left\{y(t)^6\right\}$  detailed in \eqref{y_DC_2_2}, \eqref{y_DC_4_3} and \eqref{y_DC_6_1}, respectively (at the top of next page). There is no odd (first, third, fifth, etc) order terms since $\mathcal{E}\left\{y(t)^i\right\}=\mathcal{E}\left\{y(t)^i\right\}=0$ for $i$ odd. In \eqref{yt_2} and \eqref{yt_4}, only terms with an equal number of $+$ and $-$ lead to a DC component in \eqref{y_DC_2_2} and \eqref{y_DC_4_3} following the assumption on evenly spaced frequencies. 
\par We note that the second order term \eqref{y_DC_2_2} is linear, with the DC power being the sum of the power harvested on each frequency. On the other hand, even terms with $i\geq 4$ such as \eqref{y_DC_4_3} and \eqref{y_DC_6_1} are responsible for the non-linear behaviour of the diode since they are function of terms expressed as the product of contributions from different frequencies. 

\begin{table*}
\begin{align}
\mathcal{E}\left\{y(t)^2\right\}&=\frac{1}{2}\left[\sum_{n=0}^{N-1}X_n^2\right]=\frac{1}{2}\left[\sum_{n=0}^{N-1} \left|\mathbf{h}_n\mathbf{w}_{n} \right|^2\right]=\frac{1}{2}\left[\sum_{n=0}^{N-1} \sum_{m_0,m_1} s_{n,m_0}s_{n,m_1}A_{n,m_0}A_{n,m_1}\cos\left(\psi_{n,m_0}-\psi_{n,m_1}\right)\right],\label{y_DC_2_2}\\
\mathcal{E}\left\{y(t)^4\right\}
&=\frac{3}{8}\left[\sum_{\mycom{n_0,n_1,n_2,n_3}{n_0+n_1=n_2+n_3}}X_{n_0}X_{n_1}X_{n_2}X_{n_3}\cos(\delta_{n_0}+\delta_{n_1}-\delta_{n_2}-\delta_{n_3})\right]=\frac{3}{8}\Re\left\{\sum_{\mycom{n_0,n_1,n_2,n_3}{n_0+n_1=n_2+n_3}}\mathbf{h}_{n_0}\mathbf{w}_{n_0}\mathbf{h}_{n_1}\mathbf{w}_{n_1}\left(\mathbf{h}_{n_2}\mathbf{w}_{n_2}\right)^*\left(\mathbf{h}_{n_3}\mathbf{w}_{n_3}\right)^*\right\},\label{y_DC_4_1b}\\
&=\frac{3}{8}\left[\sum_{\mycom{n_0,n_1,n_2,n_3}{n_0+n_1=n_2+n_3}}\sum_{\mycom{m_0,m_1,}{m_2,m_3}}\Bigg[\prod_{j=0}^3s_{n_j,m_j}A_{n_j,m_j}\Bigg]\cos(\psi_{n_0,m_0}+\psi_{n_1,m_1}-\psi_{n_2,m_2}-\psi_{n_3,m_3})\right].\label{y_DC_4_3}\\
\mathcal{E}\left\{y(t)^6\right\}&=\frac{5}{16}\Re\left\{\sum_{\mycom{n_0,n_1,n_2,n_3,n_4,n_5}{n_0+n_1+n_2=n_3+n_4+n_5}}\mathbf{h}_{n_0}\mathbf{w}_{n_0}\mathbf{h}_{n_1}\mathbf{w}_{n_1}\mathbf{h}_{n_2}\mathbf{w}_{n_2}\left(\mathbf{h}_{n_3}\mathbf{w}_{n_3}\right)^*\left(\mathbf{h}_{n_4}\mathbf{w}_{n_4}\right)^*\left(\mathbf{h}_{n_5}\mathbf{w}_{n_5}\right)^*\right\},\label{y_DC_6_0}\\
&=\frac{5}{16}\left[\sum_{\mycom{n_0,n_1,n_2,n_3,n_4,n_5}{n_0+n_1+n_2=n_3+n_4+n_5}}\sum_{\mycom{m_0,m_1,m_2,}{m_3,m_4,m_5}}\Bigg[\prod_{j=0}^5 s_{n_j,m_j}A_{n_j,m_j}\Bigg]\cos(\psi_{n_0,m_0}+\psi_{n_1,m_1}+\psi_{n_2,m_2}-\psi_{n_3,m_3}-\psi_{n_4,m_4}-\psi_{n_5,m_5})\right].\label{y_DC_6_1}
\end{align}\hrulefill
\end{table*}

\subsection{A Linear Model}
The linear model was first introduced a few decades ago in \cite{Wetenkamp:1983} and recently became popular in the SWIPT literature \cite{Zhou:2013}. It could be argued that if $y(t)$ is very small (i.e.\ for a very low input power), the high order ($> 2$) terms would not contribute much to $i_{out}$. Hence, the linear model truncates the Taylor expansion to the second order $n_o=2$ such that $i_{out}\approx k_0+k_2 R_{ant}\mathcal{E}\left\{y(t)^2\right\}$. It therefore completely omits the non-linearity behavior of the rectifier. The linear model is motivated by its simplicity rather than its accuracy. Its accuracy is actually questionable in the RF literature with experiments demonstrating that the non-linearity is an essential property of the rectenna and that a second order truncation of the Taylor expansion does not accurately model the rectification behavior of the diode \cite{Ladan:2015}. Nevertheless, the loss incurred by using a linear vs a non-linear model in the WPT waveform and system design has never been addressed so far.
\par In the next section, we derive tools to design waveforms under the assumption of a linear and non-linear model. 

\section{WPT Waveform Optimization}\label{section_waveform}
Assuming the CSI (in the form of frequency response $h_{n,m}$) is known to the transmitter, we aim at finding the optimal set of amplitudes and phases $\mathbf{S},\mathbf{\Phi}$ that maximizes $i_{out}$, i.e.\
\begin{equation}\label{P1}
\max_{\mathbf{S},\mathbf{\Phi}} \hspace{0.2cm} i_{out}(\mathbf{S},\mathbf{\Phi}) \hspace{0.3cm} \textnormal{subject to} \hspace{0.3cm} \frac{1}{2}\left\|\mathbf{S}\right\|_F^2\leq P.
\end{equation}
From the previous section, we however note that the rectifier characteristics ${k_i}$ are functions of $a$. Since we chose $a=-v_{out}=-R_L i_{out}$ in the Taylor expansion, $k_i$ are therefore a function of the output DC current. Making this dependence explicit, we can write $i_{out}$ from \eqref{diode_model_2} as
\begin{equation}\label{i_out_dependence} 
i_{out}\approx k_0\left(i_{out}\right)+\sum_{i \hspace{0.1cm}\textnormal{even}, i\geq 2}^{n_o} k_i\left(i_{out}\right) R_{ant}^{i/2} \mathcal{E}\left\{y(t)^i\right\}.
\end{equation}
Soving Problem \eqref{P1} with $i_{out}$ given in \eqref{i_out_dependence} may seem challenging because of the occurence of $i_{out}$ on both sides of the equality in \eqref{i_out_dependence}. Denoting $k_0'=e^{\frac{a}{nv_t}}=e^{\frac{-R_L i_{out}}{nv_t}}$ and $k_0=i_s \left(k_0'-1\right)$, we write \eqref{i_out_dependence} equivalently as
\begin{align}
e^{\frac{R_L i_{out}}{n v_t}}\left(i_{out}+i_s\right)
\approx i_s+\sum_{i \hspace{0.1cm}\textnormal{even}, i\geq 2}^{n_o} \frac{k_i}{k_0'} R_{ant}^{i/2} \mathcal{E}\left\{y(t)^i\right\}.\label{f_iout}
\end{align}
Interestingly, this leads to an expression where the r.h.s of \eqref{f_iout} is independent of $a$ (and $i_{out}$) since $k_i/k_0'=\frac{i_s}{i!\left(n v_t\right)^i}$, with $i$ even and $i\geq 2$. The l.h.s of \eqref{f_iout} is on the other hand a monotonic increasing function of $i_{out}$. Hence the maximization of $i_{out}$ is equivalent to maximizing the r.h.s of \eqref{f_iout}, which is equivalent to maximizing the quantity
\begin{equation}\label{z_DC_def}
z_{DC}(\mathbf{S},\mathbf{\Phi})=\sum_{i \hspace{0.1cm}\textnormal{even}, i\geq 2}^{n_o} k_i R_{ant}^{i/2} \mathcal{E}\left\{y(t)^i\right\}
\end{equation}  
since $i_s$ is a constant. In \eqref{z_DC_def}, we define $k_i=\frac{i_s}{i!\left(n v_t\right)^i}$ with a slight abuse of notation. Assuming $i_s=5 \mu A$, a diode ideality factor $n=1.05$ and $v_t=25.86 mV$, typical values of those parameters for second and fourth order are given by $k_2=0.0034$ and $k_4=0.3829$ (and will be used as such in any evaluation in the sequel). 
Hence, $\max_{\mathbf{S},\mathbf{\Phi}}i_{out}$ is equivalent to $\max_{\mathbf{S},\mathbf{\Phi}}z_{DC}$ and problem \eqref{P1} can equivalently be written as 
\begin{equation}\label{P2}
\max_{\mathbf{S},\mathbf{\Phi}} \hspace{0.2cm} z_{DC}(\mathbf{S},\mathbf{\Phi}) \hspace{0.3cm} \textnormal{subject to} \hspace{0.3cm} \frac{1}{2}\left\|\mathbf{S}\right\|_F^2\leq P.
\end{equation}


\subsection{Linear Model-based Design}\label{ASS}
\par With the linear model, problem \eqref{P2} is equivalent to
\begin{align}\label{P0}
\max_{\mathbf{w}_{n}} \hspace{0.3cm} \sum_{n=0}^{N-1} \left|\mathbf{h}_n\mathbf{w}_{n} \right|^2 \hspace{0.3cm}\textnormal{s.t.} \hspace{0.3cm} \frac{1}{2}\left[\sum_{n=0}^{N-1} \left\|\mathbf{w}_{n} \right\|^2\right]\leq P.
\end{align}
The solution simply consists in performing a matched beamformer on a single sinewave, namely the one corresponding to the strongest channel $\bar{n}=\arg \max_{i}\left\|\mathbf{h}_i\right\|^2$. Hence,
\begin{equation}\label{solution_2nd_order}
\mathbf{w}^{\star}_{n}=\left\{\begin{array}{l}
\sqrt{2P}\:\mathbf{h}_n^H/\left\|\mathbf{h}_n\right\|, \hspace{0.2cm} n=\bar{n}, \\
\mathbf{0}, \hspace{0.2cm} n\neq\bar{n}. 
\end{array}
\right.
\end{equation}
We denote solution \eqref{solution_2nd_order} as the adaptive single sinewave (ASS) strategy. With such a linear model, a single-sine waveform is favoured over a multisine waveform. Such a strategy has also appeared in the SWIPT literature with OFDM transmission, e.g.\ \cite{Huang:2013,Bayani:2015}.

\begin{remark} 
For the extreme case where the channel is perfectly flat magnitude-wise, i.e.\ $\left\|\mathbf{h}_n\right\|=\left\|\mathbf{h}\right\|$ $\forall n$, ASS is not the only solution to problem \eqref{P0}. Allocating power uniformly over any non-empty subset $S$ of the $N$ sinewaves, i.e.\ 
\begin{equation}\label{solution_2nd_order_flat}
\mathbf{w}^{\star}_{n}=\left\{\begin{array}{l}
\sqrt{\frac{2P}{\left\vert{S}\right\vert}}\:\mathbf{h}_n^H/\left\|\mathbf{h}\right\|, \hspace{0.2cm} n \in S, \\
\mathbf{0}, \hspace{0.2cm} n\notin S, 
\end{array}
\right.
\end{equation}
leads to the same objective value. If the channel is not perfectly flat, ASS would be the unique solution to problem \eqref{P0}.
\end{remark}

\subsection{Towards a Non-Linear Model-based Design}\label{Lagrangian_N2}
\par To get some insights into the necessity to account for the non-linear terms (e.g. $4^{th}$,$6^{th}$) and into the impact of multipath on the waveform design, let us consider a toy example with the simplest multisine: $N=2$, $M=1$. We also assume $n_o=4$. For readibility, we drop the antenna index and assume real frequency domain channel $h_{n}$. Since $\bar{\psi}_{n}=0$, let us choose $\phi_{n}=0$ so that $\psi_{n}=0$ (and all $\cos(.)=1$) in \eqref{y_DC_2_2} and \eqref{y_DC_4_3} $\forall n,m$. Since for $N=2$, indices $n_0$, $n_1$, $n_2$, $n_3$ in \eqref{y_DC_4_3} can take either value 0 or 1, we can easily identify cases for which $n_0+n_1=n_2+n_3$ and then write from \eqref{z_DC_def}
\begin{multline}
z_{DC}(s_0,s_1)=\tilde{k}_2 \left(s_0^2A_0^2+s_1^2A_1^2\right) \\
+\tilde{k}_4  \left[\left(s_0^2A_0^2+s_1^2A_1^2\right)^2+ 2 s_0^2s_1^2A_0^2A_1^2\right]\label{diode_model_2_N2_1}
\end{multline}
where $\tilde{k}_2=k_2 R_{ant}/2$ and $\tilde{k}_4=3 k_4 R_{ant}^2/8$. From \eqref{diode_model_2_N2_1}, we note that $z_{DC}(s_0,s_1)$ is a function of the term $s_0^2A_0^2+s_1^2A_1^2$, whose maximization subject to the sum power constraint $s_0^2+s_1^2\leq 2P$ would lead to the ASS strategy \eqref{solution_2nd_order}, i.e.\ allocate all the power to sinewave 1 if $A_1>A_0$ and to sinewave 0 otherwise. However the presence of the term $2 s_0^2s_1^2A_0^2A_1^2$ suggests that such a single-sinewave strategy is in general sub-optimal for the maximization of $z_{DC}$. In  problem \eqref{P2} with $N=2$ and $M=1$, we note that equality $\frac{1}{2}\left\|\mathbf{S}\right\|_F^2 = P$ is satisfied at optimality and we write the Lagrangian as
\begin{multline}
\mathcal{L}=\tilde{k}_2 \left(s_0^2A_0^2+s_1^2A_1^2\right) +\tilde{k}_4  \left(s_0^4 A_0^4+s_1^4 A_1^4+4 s_0^2s_1^2A_0^2A_1^2\right) \\ +\lambda\left(s_0^2+s_1^2-2P\right).
\end{multline}
Differentiating w.r.t.\ $s_0$, $s_1$, $\lambda$ and equating to 0, we find three valid stationary points $(s_0^2,s_1^2)$ (such that $0 \leq s_0^2\leq 2P$ and $0 \leq s_1^2\leq 2P$) given by
$(2P,0)$, $(0,2P)$ and $(s_0^{\star 2},s_1^{\star 2})$ where
\begin{align}
s_0^{\star 2}&=\frac{8P\tilde{k}_4A_0^2A_1^2+\tilde{k}_2A_0^2-4P\tilde{k}_4 A_1^4-\tilde{k}_2A_1^2}{8\tilde{k}_4A_0^2A_1^2-2\tilde{k}_4A_0^4-2\tilde{k}_4A_1^4},\\
s_1^{\star 2}&=2P-s_0^{\star 2}.
\end{align}
For given $A_0$, $A_1$, the global optimum strategy is given by one of those three stationary points. The maximum achievable $z_{DC}^{\star}=\max\big\{z_{DC}(\sqrt{2P},0),z_{DC}(0,\sqrt{2P}),z_{DC}(s_0^{\star},s_1^{\star})\big\}$. The first two points correspond to the ASS strategy, allocating transmit power to sinewave 0 and 1, respectively. Fig \ref{Lagrangian_N2} illustrates $z_{DC}$ as a function of $A_1$ for $A_0=1$ with three strategies: single-sinewave transmission (i.e.\ $s_0=0$ and $s_1=0$) and the optimal transmission leading to $z_{DC}^{\star}$. The contours of $z_{DC}$ as a function of $s_0^2$ and $s_1^2$ are also illustrated for $A_0=1$ and $A_1=0.75,1,1.15$. We note that the ASS strategy is optimal if $A_0$ is sufficiently larger than $A_1$ or inversely. However, when the channel is frequency flat, i.e.\ $A_1\approx A_0$, the optimal strategy would allocate power to the two sinewaves and the ASS strategy is suboptimal. 
  
\begin{figure}
\centerline{\includegraphics[width=0.9\columnwidth]{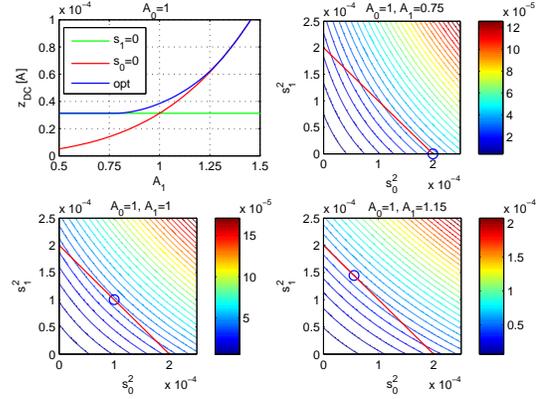}}
  \caption{$z_{DC}$ as a function of $A_1$ and contours of $z_{DC}$ as a function of $s_0^2$ and $s_1^2$. The straight line refers to the power constraint and the circle to the optimal power allocation strategy. $P=-40dBW$.}
  \label{Lagrangian_N2}
\end{figure}
The results, though based on a very simple scenario, highlight that depending on the CSI, the transmission waveform should be adapted if we aim at maximizing the output DC power. Moreover, it also shows the benefits of allocating power over multiple sinewaves for some channel states, which is in sharp contrast with the ASS strategy \eqref{solution_2nd_order} originating from the linear model. More generally, looking at \eqref{y_DC_4_3}, the ASS strategy would unlikely be a right strategy if we account for the non-linearity of the diode, due to the presence of $\prod_{j=0}^3s_{n_j,m_j}A_{n_j,m_j}$ in the fourth order term. 

\begin{remark} 
It should be noted that RF experiments in \cite{Trotter:2009,Boaventura:2011,Collado:2014} have shown the benefits of allocating power uniformly across multiple sinewaves. The above discussion highlights theoretically the benefits of allocating power over multiple sinewaves for some channel states and therefore backs up the experimental results. On the other hand, the linear model motivates the use of a single sinewave (ASS) for all channel states, and therefore contradicts the RF experiment results. 
\end{remark}

\par Deriving a formal algorithm that can generate optimized waveforms for any multipath configuration and any $N$, $M$, $n_o$ so as to maximize the DC output current is a non-trivial problem that is discussed in the next section.

\subsection{Non-Linear Model-based Design}\label{nonlinear_design_subsection}
\par We aim at deriving a waveform design strategy that is general enough to cope with any Taylor expansion order $n_o$\footnote{We display terms for $n_o\leq 6$ but the derived algorithm works for any $n_o$.}. The optimal phases $\mathbf{\Phi}$ can be obtained first in closed form and the optimal amplitudes $\mathbf{S}$ can then be computed numerically.

\par To maximize $z_{DC}(\mathbf{S},\mathbf{\Phi})$, we should guaranteee all $\cos(.)$ to be equal to 1 in \eqref{y_DC_2_2}, \eqref{y_DC_4_3} and \eqref{y_DC_6_1}. This can be satisfied by choosing $\psi_{n,m}=0$ $\forall n,m$ (and therefore $\delta_{n}=0$ $\forall n$), which implies from \eqref{A_nm_psi} to choose the optimal sinewave phases as
\begin{equation}\label{opt_phases}
\phi_{n,m}^{\star}=-\bar{\psi}_{n,m}.
\end{equation}
$\mathbf{\Phi}^{\star}$ is obtained by collecting $\phi_{n,m}^{\star}$ $\forall n,m$ into matrix. With such a phase choice, all sinewaves in \eqref{y_t} are in-phase at the rectenna input. Moreover, $\psi_{n,m}=0$ and $X_n=\sum_{m=1}^M s_{n,m}A_{n,m}$ such that $z_{DC}(\mathbf{S},\mathbf{\Phi}^{\star})$ is simply obtained from \eqref{z_DC_def} with all $\cos(.)$ replaced by 1 in \eqref{y_DC_2_2}, \eqref{y_DC_4_3} and \eqref{y_DC_6_1}.
\par Recall from \cite{Chiang:2005} that a monomial is defined as the function $g:\mathbb{R}_{++}^{N}\rightarrow\mathbb{R}:g(\mathbf{x})=c x_1^{a_1}x_2^{a_2}\ldots x_N^{a_N}$ 
where $c>0$ and $a_i\in\mathbb{R}$. A sum of $K$ monomials is called a posynomial and can be written as 
$f(\mathbf{x})=\sum_{k=1}^K g_k(\mathbf{x})$ with $g_k(\mathbf{x})=c_k x_1^{a_{1k}}x_2^{a_{2k}}\ldots x_N^{a_{Nk}}$
where $c_k>0$. As we can see from \eqref{y_DC_2_2}, \eqref{y_DC_4_3} and \eqref{y_DC_6_1}, $z_{DC}(\mathbf{S},\mathbf{\Phi}^{\star})$ is a posynomial, and so it is  
for any order $n_o$ in the Taylor expansion. The higher the order, the larger the number of terms in the posynomial.

\par The optimization problem becomes $\max_{\mathbf{S}} z_{DC}(\mathbf{S},\mathbf{\Phi}^{\star})$ subject to $\frac{1}{2}\left\|\mathbf{S}\right\|_F^2\leq P$. It therefore consists in maximizing a posynomial subject to a power constraint (which itself is written as a posynomial). 
This problem is not a standard Geometric Program (GP) but it can be transformed to an equivalent problem by introducing an auxiliary variable $t_0$
\begin{align}\label{reverse_GP}
\min_{\mathbf{S},t_0} \hspace{0.3cm} &1/t_0\\
\textnormal{subject to} \hspace{0.3cm} &\frac{1}{2}\left\|\mathbf{S}\right\|_F^2\leq P,\\ 
&z_{DC}(\mathbf{S},\mathbf{\Phi}^{\star})/t_0\geq 1.\label{reverse_GP_2}
\end{align}

\par This is known as a Reverse Geometric Program due to the minimization of a posynomial subject to upper and lower bounds inequality constraints \cite{Duffin:1973,Chiang:2005}. Note that  $z_{DC}(\mathbf{S},\mathbf{\Phi}^{\star})/t_0\geq 1$ is equivalent to $t_0/z_{DC}(\mathbf{S},\mathbf{\Phi}^{\star})\leq1$. However $1/z_{DC}(\mathbf{S},\mathbf{\Phi}^{\star})$ is not a posynomial, therefore preventing the use of standard GP tools.
The idea is to lower bound $z_{DC}(\mathbf{S},\mathbf{\Phi}^{\star})$ by a monomial $\bar{z}_{DC}(\mathbf{S})$, i.e.\ upper bound $1/z_{DC}(\mathbf{S},\mathbf{\Phi}^{\star})$ by the monomial $1/\bar{z}_{DC}(\mathbf{S})$ (since the inverse of a monomial is still a monomial) \cite{Duffin:1973}. Let $\left\{g_k(\mathbf{S},\mathbf{\Phi}^{\star})\right\}$ be the monomial terms in the posynomial $z_{DC}(\mathbf{S},\mathbf{\Phi}^{\star})=\sum_{k=1}^K g_k(\mathbf{S},\mathbf{\Phi}^{\star})$. The choice of the lower bound relies on the fact that an arithmetic mean (AM) is greater or equal to the geometric mean (GM). Hence, $z_{DC}(\mathbf{S},\mathbf{\Phi}^{\star})\geq \prod_{k=1}^K\left(\frac{g_k(\mathbf{S},\mathbf{\Phi}^{\star})}{\gamma_k}\right)^{\gamma_k}=\bar{z}_{DC}(\mathbf{S})$, where $\gamma_k\geq 0$ and $\sum_{k=1}^K \gamma_k=1$. Since
\begin{equation}\label{UB}
1/z_{DC}(\mathbf{S},\mathbf{\Phi}^{\star})\leq 1/\bar{z}_{DC}(\mathbf{S}),
\end{equation}
we can replace (in a conservative way) inequality $t_0/z_{DC}(\mathbf{S},\mathbf{\Phi}^{\star})\leq1$ by $t_0/\bar{z}_{DC}(\mathbf{S})=t_0\prod_{k=1}^K\left(g_k(\mathbf{S},\mathbf{\Phi}^{\star})/\gamma_k\right)^{-\gamma_k}\leq 1$. 
For a given choice of $\left\{\gamma_k\right\}$, problem \eqref{reverse_GP}-\eqref{reverse_GP_2} is now replaced by the standard GP
\begin{align}\label{standard_GP}
\min_{\mathbf{S},t_0} \hspace{0.3cm} &1/t_0\\
\textnormal{subject to} \hspace{0.3cm} &\frac{1}{2}\left\|\mathbf{S}\right\|_F^2\leq P,\\ 
&t_0\prod_{k=1}^K\left(\frac{g_k(\mathbf{S},\mathbf{\Phi}^{\star})}{\gamma_k}\right)^{-\gamma_k}\leq1,\label{standard_GP_constraint_2}
\end{align}
that can be solved using existing software, e.g.\ CVX \cite{CVX}.

\par Note that the tightness of the upper bound \eqref{UB} heavily depends on the choice of $\left\{\gamma_k\right\}$. Following \cite{Beightler:1976,Chiang:2005}, an iterative procedure can be used to tighten the bound, where at each iteration the standard GP \eqref{standard_GP}-\eqref{standard_GP_constraint_2} is solved for an updated set of $\left\{\gamma_k\right\}$. Assuming a feasible set of magnitude $\mathbf{S}^{(i-1)}$ at iteration $i-1$, compute at iteration $i$  $\gamma_k=g_k(\mathbf{S}^{(i-1)},\mathbf{\Phi}^{\star})/z_{DC}(\mathbf{S}^{(i-1)},\mathbf{\Phi}^{\star})$ $\forall k$ and solve problem \eqref{standard_GP}-\eqref{standard_GP_constraint_2} to obtain $\mathbf{S}^{(i)}$. Repeat the iterations till convergence. Algorithm \ref{Algthm_OPT} summarizes the procedure.

\begin{algorithm}
\caption{WPT Waveform}
\label{Algthm_OPT}
\begin{algorithmic}[1]
\State \textbf{Initialize}: $i\gets 0$, $\mathbf{\Phi}^{\star}$ in \eqref{opt_phases}, $\mathbf{S}$, $z_{DC}^{(0)}=0$
\label{Algthm_OPT_step_initialize}
\Repeat
    \State $i\gets i+1$, $\ddot{\mathbf{S}}\gets \mathbf{S}$
    \State $\gamma_k\gets g_k(\ddot{\mathbf{S}},\mathbf{\Phi}^{\star})/z_{DC}(\ddot{\mathbf{S}},\mathbf{\Phi}^{\star})$, $k=1,\ldots,K$  
    \label{Algthm_OPT_step_gamma}
    \State  $\mathbf{S} \gets \arg \min \eqref{standard_GP}-\eqref{standard_GP_constraint_2}$
    \label{Algthm_OPT_step_OPT}
    \State $z_{DC}^{(i)} \gets z_{DC}(\mathbf{S},\mathbf{\Phi}^{\star})$
\Until{$\left|z_{DC}^{(i)} - z_{DC}^{(i-1)} \right| < \epsilon$ \text{or} $i=i_{\max}$ }
\end{algorithmic}
\end{algorithm}

Note that the successive approximation method used in the Algorithm 1 is also known as a sequential convex optimization or inner approximation method \cite{Marks:1978}. It cannot guarantee to converge to the global solution of the original problem, but only to yield a point fulfilling the KKT conditions \cite{Marks:1978,Chiang:2007}. However, it has been shown in \cite{Chiang:2005} by simulation that such an iterative algorithm often converges to the global optimum.

\begin{remark} Non-linearity is obviously meaningful only for $N\geq 2$. For $N=1$, both linear and non-linear designs boil down to the simple matched beamformer $\mathbf{w}=\sqrt{2P}\mathbf{h}^H/\left\|\mathbf{h}\right\|$.
\end{remark}

\subsection{Decoupling Space and Frequency Domains}\label{decoupling}
When $M>1$, previous section derives a general methodology to design waveform weights jointly across space and frequency. It is worth wondering whether we can decouple the design of the spatial and frequency domain weights without impacting performance. The optimal phases in \eqref{opt_phases} are those of a matched beamformer. Looking at \eqref{y_DC_2_2}, \eqref{y_DC_4_1b} and \eqref{y_DC_6_0}, the optimum weight vector $\mathbf{w}_n$ that maximizes the $2^{nd}$, $4^{th}$ and $6^{th}$ order terms is actually a matched beamformer of the form
\begin{equation}\label{opt_weight}
\mathbf{w}_n=s_n \mathbf{h}_n^H/\left\|\mathbf{h}_n\right\| 
\end{equation} 
such that, from \eqref{y_t}, $y(t)=\sum_{n=0}^{N-1}\left\|\mathbf{h}_n\right\| s_n \cos\left(w_n t\right)=\Re\left\{\sum_{n=0}^{N-1}\left\|\mathbf{h}_n\right\| s_n e^{j w_n t}\right\}$.
Hence, with \eqref{opt_weight}, the multi-antenna multi-sine WPT weight optimization is converted into an effective single antenna multi-sine WPT optimization with the effective channel gain on frequency $n$ given by $\left\|\mathbf{h}_n\right\|$ and the amplitude of the $n^{th}$ sinewave given by $s_n$ (subject to $\sum_{n=0}^{N-1}s_n^2=2P$). The optimum magnitude $s_n$ in \eqref{opt_weight} can now be obtained by using the posynomial maximization methodology of Section \ref{nonlinear_design_subsection}. Namely, plugging \eqref{opt_weight} into \eqref{y_DC_2_2}, \eqref{y_DC_4_1b} and \eqref{y_DC_6_0}, we get \eqref{y_DC_w_3}.
\begin{table*}
\begin{align}
\mathcal{E}\left\{y(t)^2\right\}=\frac{1}{2}\left[\sum_{n=0}^{N-1} \left\|\mathbf{h}_n\right\|^2 s_n^2\right],
\mathcal{E}\left\{y(t)^4\right\}=\frac{3}{8}\left[\sum_{\mycom{n_0,n_1,n_2,n_3}{n_0+n_1=n_2+n_3}}\Bigg[\prod_{j=0}^3s_{n_j}\left\|\mathbf{h}_{n_j}\right\|\Bigg]\right],
\mathcal{E}\left\{y(t)^6\right\}=\frac{5}{16}\left[\sum_{\mycom{n_0,n_1,n_2,n_3,n_4,n_5}{n_0+n_1+n_2=n_3+n_4+n_5}}\Bigg[\prod_{j=0}^5 s_{n_j}\left\|\mathbf{h}_{n_j}\right\|\Bigg]\right]\label{y_DC_w_3}
\end{align}\hrulefill
\end{table*}
$z_{DC}\left(\mathbf{s}\right)=\sum_{i \hspace{0.1cm}\textnormal{even}, i\geq 2}^{n_o} k_i R_{ant}^{i/2} \mathcal{E}\left\{y(t)^i\right\}$ is now only a function of the $N$-dimensional vector $\mathbf{s}=\left[\begin{array}{c} s_0,\ldots,s_{N-1}\end{array}\right]$. Following the posynomial maximization methodology, we can write $z_{DC}(\mathbf{s})=\sum_{k=1}^K g_k(\mathbf{s})$, apply the AM-GM inequality and write the standard GP problem
\begin{align}\label{standard_GP_simple}
\min_{\mathbf{s},t_0} \hspace{0.3cm} &1/t_0\\
\textnormal{subject to} \hspace{0.3cm} &\frac{1}{2}\left\|\mathbf{s}\right\|^2\leq P,\\ 
&t_0\prod_{k=1}^K\left(\frac{g_k(\mathbf{s})}{\gamma_k}\right)^{-\gamma_k}\leq1.\label{standard_GP_constraint_2_simple}
\end{align}
Algorithm \ref{Algthm_OPT_simple} summarizes the design methodology with spatial and frequency domain decoupling.
\begin{algorithm}
\caption{WPT Waveform with Decoupling}
\label{Algthm_OPT_simple}
\begin{algorithmic}[1]
\State \textbf{Initialize}: $i\gets 0$, $\mathbf{w}_n$ in \eqref{opt_weight}, $\mathbf{s}$, $z_{DC}^{(0)}=0$
\label{Algthm_OPT_step_initialize_simple}
\Repeat
    \State $i\gets i+1$, $\ddot{\mathbf{s}}\gets \mathbf{s}$
    \State $\gamma_k\gets g_k(\ddot{\mathbf{s}})/z_{DC}(\ddot{\mathbf{s}})$, $k=1,\ldots,K$  
    \label{Algthm_OPT_step_gamma_simple}
    \State  $\mathbf{s} \gets \arg \min \eqref{standard_GP_simple}-\eqref{standard_GP_constraint_2_simple}$
    \label{Algthm_OPT_step_OPT_simple}
    \State $z_{DC}^{(i)} \gets z_{DC}(\mathbf{s})$
\Until{$\left|z_{DC}^{(i)} - z_{DC}^{(i-1)} \right| < \epsilon$ \text{or} $i=i_{\max}$ }
\end{algorithmic}
\end{algorithm}
Such an approach would lead to the same performance as the joint space-frequency design of Algorithm \ref{Algthm_OPT} but would significantly reduce the computational complexity since only a $N$-dimensional vector $\mathbf{s}$ is to be optimized numerically, compared to the $N\times M$ matrix $\mathbf{S}$ of Algorithm \ref{Algthm_OPT}.
  
\subsection{PAPR Constraints}
In practice, it may be useful to constrain the PAPR of the transmitted waveform in order to increase the efficiency of the power amplifier. From \eqref{WPT_waveform}, the PAPR on antenna $m$ can be defined as
$PAPR_m=\frac{\max_t\left|x_m(t)\right|^2}{\mathcal{E}\small\{|x_m(t)|^2\small\}}=\frac{\max_t\left|x_m(t)\right|^2}{\frac{1}{2}\left\|\mathbf{s}_m\right\|^2}$.
The PAPR constraint on antenna $m$ writes as $PAPR_m\leq \eta$. Problem \eqref{P1} is now subject to an extra constraint
\begin{align}\label{P2_P}
\max_{\mathbf{S},\mathbf{\Phi}} \hspace{0.3cm} &i_{out}(\mathbf{S},\mathbf{\Phi})\\
\textnormal{subject to} \hspace{0.3cm} &\frac{1}{2}\left\|\mathbf{S}\right\|_F^2\leq P,\\
&PAPR_m\leq \eta, \forall m.\label{P2_2}
\end{align}
In the sequel, we will assume the use of the phase $\mathbf{\Phi}^{\star}$ in \eqref{opt_phases} and optimize the amplitude $\mathbf{S}$.
\par By oversampling the transmit signals at $t_q=q\frac{T}{N O_s}$ for $q=0,\ldots,N O_s-1$ with $T=1/\Delta_f=\frac{2\pi}{\Delta_w}$ and $O_s$ the oversampling factor, the PAPR constraint can be rewritten as $\left|x_m(t_q)\right|^2 \leq\frac{1}{2} \eta \left\|\mathbf{s}_m\right\|^2$, $\forall q=0,\ldots,N O_s-1$
for sufficiently large $O_s$.
Assuming the phase $\mathbf{\Phi}^{\star}$ in \eqref{opt_phases}, we can write 
\begin{multline}
\left|x_m(t_q)\right|^2=\sum_{n_0,n_1}s_{n_0,m}s_{n_1,m}
\\\cos\left(w_{n_0} t_q+\phi^{\star}_{n_0,m}\right)\cos\left(w_{n_1} t_q+\phi^{\star}_{n_1,m}\right).
\end{multline}
The quantity $\left|x_m(t_q)\right|^2$ is not a posynomial anymore as some of the coefficients $c_k$ are negative. $\left|x_m(t_q)\right|^2$ is now written as a signomial, i.e.\ the sum of monomials whose coefficients $c_k$ can be either positive or negative,
$f(\mathbf{x})=f_1(\mathbf{x})-f_2(\mathbf{x})$
where $f_j(\mathbf{x})=\sum_{k=1}^{K_j} g_{jk}(\mathbf{x})$ and $g_{jk}(\mathbf{x})=c_{jk} x_1^{a_{1jk}}x_2^{a_{2jk}}\ldots x_N^{a_{Njk}}$ with $c_{jk}>0$. Let us write the signomial $\left|x_m(t_q)\right|^2=f_{mq}(\mathbf{S},\mathbf{\Phi}^{\star})=f_{mq1}(\mathbf{S},\mathbf{\Phi}^{\star})-f_{mq2}(\mathbf{S},\mathbf{\Phi}^{\star})$. We therefore have the inequality $f_{mq1}(\mathbf{S},\mathbf{\Phi}^{\star})-f_{mq2}(\mathbf{S},\mathbf{\Phi}^{\star})\leq \frac{1}{2} \eta \left\|\mathbf{s}_m\right\|^2$ or equivalently $\frac{f_{mq1}(\mathbf{S},\mathbf{\Phi}^{\star})}{\frac{1}{2} \eta \left\|\mathbf{s}_m\right\|^2+f_{mq2}(\mathbf{S},\mathbf{\Phi}^{\star})}\leq 1$. 
This is a standard sign inequality but the quotient of posynomials is not a posynomial. Writing the denominator as a sum of monomials, $\frac{1}{2} \eta \left\|\mathbf{s}_m\right\|^2+f_{mq2}(\mathbf{S},\mathbf{\Phi}^{\star})=\sum_{k=1}^{K_{mq2}}g_{mq2k}(\mathbf{S},\mathbf{\Phi}^{\star})$, we can perform a single condensation and replace the original inequality by the following inequality
\begin{equation}
f_{mq1}(\mathbf{S},\mathbf{\Phi}^{\star}) \prod_{k=1}^{K_{mq2}}\left(\frac{g_{mq2k}(\mathbf{S},\mathbf{\Phi}^{\star})}{\gamma_{mq2k}}\right)^{-\gamma_{mq2k}}\leq 1
\end{equation}
with $\gamma_{mq2k}\geq 0$ and $\sum_{k=1}^{K_{mq2}}\gamma_{mq2k}=1$. For a given choice of $\left\{\gamma_k,\gamma_{mq2k}\right\}$ and assuming $\mathbf{\Phi}^{\star}$, the optimization problem \eqref{P2_P}-\eqref{P2_2} is now replaced by the standard GP
\begin{align}\label{standard_GP_PAPR}
\min_{\mathbf{S},t_0} \hspace{0.3cm} &1/t_0\\
\textnormal{s.t.} \hspace{0.3cm} &\frac{1}{2}\left\|\mathbf{S}\right\|_F^2\leq P,\\
&t_0\prod_{k=1}^K\left(\frac{g_k(\mathbf{S},\mathbf{\Phi}^{\star})}{\gamma_k}\right)^{-\gamma_k}\leq1,\\
&f_{mq1}(\mathbf{S},\mathbf{\Phi}^{\star}) \prod_{k=1}^{K_{mq2}}\left(\frac{g_{mq2k}}{\gamma_{mq2k}}\right)^{-\gamma_{mq2k}}\leq 1, \forall m,q \label{standard_GP_PAPR_3}
\end{align}
Problem \eqref{standard_GP_PAPR}-\eqref{standard_GP_PAPR_3} can now be solved at each iteration of an iterative procedure where $\left\{\gamma_k,\gamma_{mq2k}\right\}$ are updated. The whole optimization procedure is summarized in Algorithm \ref{Algthm_OPT_PAPR}.

\begin{algorithm}
\caption{WPT Waveform with PAPR Constraints}
\label{Algthm_OPT_PAPR}
\begin{algorithmic}[1]
\State \textbf{Initialize}: $i\gets 0$, $\mathbf{\Phi}^{\star}$ in \eqref{opt_phases}, $\mathbf{S}$, $z_{DC}^{(0)}=0$
\label{Algthm_OPT_PAPR_step_initialize}
\Repeat
    \State $i\gets i+1$, $\ddot{\mathbf{S}}\gets \mathbf{S}$
    \State $\gamma_k\gets g_k(\ddot{\mathbf{S}},\mathbf{\Phi}^{\star})/z_{DC}(\ddot{\mathbf{S}},\mathbf{\Phi}^{\star})$, $k=1,\ldots,K$  
    \label{Algthm_OPT_PAPR_step_gamma}
    \State $\gamma_{mq2k}\gets g_{mq2k}(\ddot{\mathbf{S}},\mathbf{\Phi}^{\star})/\big(\frac{1}{2} \eta \left\|\ddot{\mathbf{s}}_m\right\|^2+f_{mq2}(\ddot{\mathbf{S}},\mathbf{\Phi}^{\star})\big)$, $m=1,\ldots,M$, $q=0,\ldots,NO_s-1$, $k=1,\ldots,K_{mq2}$  
    \label{Algthm_OPT_PAPR_step_gamma_PAPR}
    \State  $\mathbf{S} \gets \arg \min \eqref{standard_GP_PAPR}-\eqref{standard_GP_PAPR_3}$
    \label{Algthm_OPT_step_OPT}
    \State $z_{DC}^{(i)} \gets z_{DC}(\mathbf{S},\mathbf{\Phi}^{\star})$
\Until{$\left|z_{DC}^{(i)} - z_{DC}^{(i-1)} \right| < \epsilon$ \text{or} $i=i_{\max}$ }
\end{algorithmic}
\end{algorithm}

Note that for $M>1$, decoupling the space and frequency domains (similarly to Section \ref{decoupling}) would lead to a suboptimal design compared to the joint space-frequency design of Algorithm \ref{Algthm_OPT_PAPR} in the presence of PAPR constraints.


%

\subsection{Multiple Rectennas}\label{MU_WPT}
Consider now the extension to $U$ rectennas. Those rectennas could either belong to a single user (i.e.\ point-to-point MIMO WPT) or spread across multiple users (i.e.\ MU-MISO WPT). In this multiple rectenna setup, the energy harvested by a given rectenna $z_{DC,q}$ in general depends on the energy harvested by the other rectennas $z_{DC,p}$, $p\neq q$. Indeed, a given waveform may be suitable for a given rectenna but found inefficient for another rectenna. Hence, there exists a trade-off between the energy harvested by the different rectennas. The energy region $\mathcal{Z}_{DC}$ formulates this trade-off by expressing the set of all rectenna harvested energy $\left(z_{DC,1},\ldots,z_{DC,U}\right)$ that are simultaneously achievable. The boundary of the energy region can be derived by considering a weighted sum of DC component at each user where weights $v_u$, $u=1,\ldots,U$, account for the multi-rectenna fairness\footnote{In the MIMO WPT, fairness among rectennas is less of an issue and a sum of DC components would be more meaningful. Weights can then simply be taken equal to 1. Another interesting architecture for the MIMO WPT (left for future studies) is such that the signals at different antennas are combined in the RF domain before being conveyed to a single rectifier.}. 

\par The optimization problem now consists in finding the optimal set of amplitudes and phases (across frequencies) that maximizes the weighted sum of DC components $z_{DC,u}$, i.e.\
\begin{equation}
\max_{\mathbf{S},\mathbf{\Phi}} Z_{DC}(\mathbf{S},\mathbf{\Phi})=\sum_{u=1}^U v_u z_{DC,u}(\mathbf{S},\mathbf{\Phi}) \hspace{0.2cm}  \textnormal{s.t.} \hspace{0.2cm} \frac{1}{2}\left\|\mathbf{S}\right\|_F^2\leq P.
\end{equation}

From Section \ref{system_model}, after adding the index $u$ to any user specific variable, we define $X_{n,u} e^{j \delta_{n,u}}=\sum_{m=1}^M s_{n,m}A_{n,m,u}e^{j \psi_{n,m,u}}$ and $A_{n,m,u}e^{j \psi_{n,m,u}}=e^{j \phi_{n,m}}h_{n,m,u}$
with $h_{n,m,u}=A_{n,m,u}e^{j \bar{\psi}_{n,m,u}}$ the frequency response of the channel of rectenna $u$ on transmit antenna $m$ at $w_n$. 

\subsubsection{Linear Model} The ASS strategy \eqref{solution_2nd_order} is generalized as 
\begin{equation}
\max_{\mathbf{w}_{n}} \hspace{0.3cm} \sum_{n=0}^{N-1} \big\|\tilde{\mathbf{H}}_{n}\mathbf{w}_{n}\big\|^2 \hspace{0.3cm}\textnormal{s.t.}\hspace{0.3cm} \frac{1}{2}\left[\sum_{n=0}^{N-1} \left\|\mathbf{w}_{n} \right\|^2\right]\leq P
\end{equation}
with $\tilde{\mathbf{H}}_{n}=\big[\begin{array}{ccc}\tilde{\mathbf{h}}_{n,1}^T & \ldots & \tilde{\mathbf{h}}_{n,U}^T \end{array}\big]^T$ and $\tilde{\mathbf{h}}_{n,u}=\sqrt{k_{2} v_u} \mathbf{h}_{n,u}$. The solution consists in transmitting on a single sinewave $\bar{n}=\arg \max_{i} \lambda_{max}\big(\tilde{\mathbf{H}}_{i}^H\tilde{\mathbf{H}}_{i}\big)$ and along the dominant right singular vector of $\tilde{\mathbf{H}}_{\bar{n}}$. Hence,
\begin{equation}\label{solution_2nd_order_MU}
\mathbf{w}^{\star}_{n}=\left\{\begin{array}{l}
\sqrt{2P}\:\mathbf{v}_{max,n}, \hspace{0.2cm} n=\bar{n}, \\
\mathbf{0}, \hspace{0.2cm} n\neq\bar{n}, 
\end{array}
\right.
\end{equation}
where $\mathbf{v}_{max,n}$ is the dominant right singular vector of $\tilde{\mathbf{H}}_{n}$. Solution \eqref{solution_2nd_order_MU} naturally boils down to \eqref{solution_2nd_order} for $U=1$.
\subsubsection{Non-Linear Model} 
Unfortunately, guaranteeing $\psi_{n,m,u}=0$ $\forall n,m,u$ is not possible ($NMU$ constraints and $NM$ variables only). 
Hence, for a given choice of phase matrix $\mathbf{\Phi}=\mathbf{\Phi}'$, some cosine functions in \eqref{y_DC_2_2}, \eqref{y_DC_4_3} and \eqref{y_DC_6_1} are positive while others are negative. $Z_{DC}(\mathbf{S},\mathbf{\Phi}')$ is now a signomial since some of the coefficients $c_k$ are negative. 

\par Similarly to the single rectenna scenario, we can convert the maximization problem into a minimization by introducing the auxiliary variable $t_0$. 
The problem writes as \eqref{reverse_GP}-\eqref{reverse_GP_2} with \eqref{reverse_GP_2} replaced by $Z_{DC}(\mathbf{S},\mathbf{\Phi}')/t_0\geq 1$.
Condition $Z_{DC}(\mathbf{S},\mathbf{\Phi}')=f_1(\mathbf{S},\mathbf{\Phi}')-f_2(\mathbf{S},\mathbf{\Phi}')\geq t_0$ can be replaced by
\begin{equation}
\frac{t_0+f_2(\mathbf{S},\mathbf{\Phi}')}{\bar{f}_1(\mathbf{S},\mathbf{\Phi}')}=\left(t_0+f_2(\mathbf{S},\mathbf{\Phi}')\right)\prod_{k=1}^{K_1}\left(\frac{g_{1k}}{\gamma_{1k}}\right)^{-\gamma_{1k}}\leq 1
\end{equation}
where $\gamma_{1k}\geq 0$, $\sum_{k=1}^{K_1} \gamma_{1k}=1$ and $\left\{g_{1k}\right\}$ are the monomial terms in the posynomial $f_1(\mathbf{S},\mathbf{\Phi}')=\sum_{k=1}^{K_1} g_{1k}(\mathbf{S},\mathbf{\Phi}')$.
For a given choice of $\left\{\gamma_{1k}\right\}$, we now get the standard GP
\begin{align}\label{standard_GP_MU_2}
\min_{\mathbf{S},t_0} \hspace{0.3cm} &1/t_0\\
\textnormal{s.t.} \hspace{0.3cm} &\frac{1}{2}\left\|\mathbf{S}\right\|_F^2\leq P,\\ 
&\left(t_0+f_2(\mathbf{S},\mathbf{\Phi}')\right)\prod_{k=1}^{K_1}\left(\frac{g_{1k}(\mathbf{S},\mathbf{\Phi}')}{\gamma_{1k}}\right)^{-\gamma_{1k}}\leq 1.\label{standard_GP_MU_constraint_4}
\end{align}
Similarly to the single rectenna optimization, Problem \eqref{standard_GP_MU_2}-\eqref{standard_GP_MU_constraint_4} can now be solved at each iteration of an iterative procedure where $\left\{\gamma_{1k}\right\}$ are updated.
Note that Problem \eqref{standard_GP_MU_2}-\eqref{standard_GP_MU_constraint_4} boils down to Problem \eqref{standard_GP}-\eqref{standard_GP_constraint_2} if $f_2=0$. The whole optimization procedure is summarized in Algorithm \ref{Algthm_OPT_MU}.

\begin{algorithm}
\caption{WPT Waveform with Multiple Rectennas}
\label{Algthm_OPT_MU}
\begin{algorithmic}[1]
\State \textbf{Initialize}: $i\gets 0$, $\mathbf{\Phi}'$, $\mathbf{S}$, $Z_{DC}^{(0)}=0$
\label{Algthm_OPT_MU_step_initialize}
\Repeat
    \State $i\gets i+1$, $\ddot{\mathbf{S}}\gets \mathbf{S}$
    \State $\gamma_{1k}\gets g_{1k}(\ddot{\mathbf{S}},\mathbf{\Phi}')/f_1(\ddot{\mathbf{S}},\mathbf{\Phi}')$, $k=1,\ldots,K_1$  
    \label{Algthm_OPT_MU_step_gamma}
    \State  $\mathbf{S} \gets \arg \min \eqref{standard_GP_MU_2}-\eqref{standard_GP_MU_constraint_4}$
    \label{Algthm_OPT_MU_step_OPT}
    \State $Z_{DC}^{(i)} \gets Z_{DC}(\mathbf{S},\mathbf{\Phi}')$
\Until{$\left|Z_{DC}^{(i)} - Z_{DC}^{(i-1)} \right| < \epsilon$ \text{or} $i=i_{\max}$ }
\end{algorithmic}
\end{algorithm}
 
\par We note that the optimum phases for the single rectenna scenario in \eqref{opt_phases} and the linear model optimization in \eqref{solution_2nd_order_MU} are those of a dominant eigenmode transmission (boiling down to a simple transmit matched filter for the single rectenna case) \cite{Clerckx:2013}. Motivated by this observation, a good choice for the phase $\mathbf{\Phi}'$ in Algorithm \ref{Algthm_OPT_MU} (even though there is no claim of optimality) consists in choosing the $\left(n,m\right)$ entries of $\mathbf{\Phi}'$ as $\phi'_{n,m}=\textnormal{phase}\left(v_{max,n,m}\right)$ where $v_{max,n,m}$ is the $m^{th}$ entry of the dominant right singular vector $\mathbf{v}_{max,n}$, $\forall n,m$.  

\subsection{CSI Acquisition at the Transmitter}\label{CSI_acq}
The proposed waveform design relies on CSI (in the form of frequency response $h_{n,m}$) knowledge at the transmitter. 
Inspired by communication systems in a TDD mode, we could envision a WPT architecture equipped with a pilot transmission (on the uplink) phase and a channel estimator at the power base station. Alternatively, approaches relying on CSI feedback, along the lines of e.g.\ \cite{Xu:2014}, could be exploited. Note that since the linear and nonlinear models give very different waveform strategies (the first one favouring a single sinewave while the second one favouring multiple sinewaves), the CSI feedback/estimation mechanisms and requirements depend on the adopted model. 
\par We may be tempted to think that the design requires knowledge of the rectifier characteristics since the parameters $k_i$ are function of $i_s$ and $v_t$. However, $i_s$ is just a mutiplicative factor affecting all terms equally in $z_{DC}$ and therefore has no impact of the design of the waveform. $v_t$ appears in the denominator of $k_i$ through the term $v_t^i$. However $v_t$ is a constant irrespective of the rectifier design. Hence the waveform design at the transmitter does not require any feedback of information about the rectifier characteristics.
 
\section{Scaling Laws of WPT} \label{section_scaling_law}
In order to further motivate the usefulness of multisine waveform optimization and in order to get some insight into the fundamental limits of WPT, we want to quantify how $z_{DC}$ and $Z_{DC}$ scale as a function of $N$, $M$ and $U$. For simplicity we truncate the Taylor expansion to the fourth order and therefore consider the metric $z_{DC}(\mathbf{S},\mathbf{\Phi})=k_2 R_{ant} \mathcal{E}\left\{y(t)^2\right\}+k_4 R_{ant}^2 \mathcal{E}\left\{y(t)^4\right\}$.
The scaling laws also draw insights into the usefulness of CSIT for WPT. We consider frequency-flat and frequency-selective channels.
\par We assume that the complex channel gains $\alpha_l e^{j\xi_l}$ are modeled as independent circularly symmetric complex Gaussian random variables. $\alpha_l$ are therefore independent Rayleigh distributed such that $\alpha_l^2\sim \textnormal{EXPO}(\lambda_l)$ with $1/\lambda_l=\beta_l=\mathcal{E}\left\{\alpha_l^2\right\}$. The impulse responses have a constant average received power normalized to 1 such that $\sum_{l=0}^{L-1}\beta_l=1$.

\subsection{Frequency-Flat Channels}
We first assume a single transmit antenna (and drop the antenna index) and a single rectenna ($U=1$) and consider a frequency flat channel with $\bar{\psi}_{n}=\bar{\psi}$ and $A_n=A$ $\forall n$. This is met when the bandwidth of the multisine waveform $(N-1)\Delta_f$ is much smaller than the channel coherence bandwidth. 

\par Making use of \eqref{y_DC_2_2}, \eqref{y_DC_4_3}, \eqref{y_DC_6_1} and \eqref{z_DC_def}, it is clear that choosing $\mathbf{\Phi}^{\star}=\mathbf{0}_{N}$ is optimal for any $A$ and $\bar{\psi}$. Recalling the power constraint $\sum_n s_n^2=2P$, we can then write 
\begin{equation}
z_{DC}\left(\mathbf{S},\mathbf{\Phi}^{\star}\right)= k_2 A^2 R_{ant} P+\frac{3 k_4}{8} A^4 R_{ant}^2 F\label{z_DC_FF_general}
\end{equation}
where 
\begin{equation}\label{F}
F=\sum_{\mycom{n_0,n_1,n_2,n_3}{n_0+n_1=n_2+n_3}}s_{n_0}s_{n_1}s_{n_2}s_{n_3}.
\end{equation}
Finding a closed form solution of the optimal $\mathbf{S}$ is challenging. We can lower bound $F$ as
$F\geq \sum_{n=0}^{N-1} s_n^4+2 \sum_{\mycom{n_0,n_1}{n_0\neq n_1}}s_{n_0}^2 s_{n_1}^2=4P^2+2 \sum_{\mycom{n_0,n_1}{n_0 < n_1}}s_{n_0}^2 s_{n_1}^2$.
Subject to the power constraint, the lower bound is maximized by allocating power uniformly across sinewaves, i.e.\ $s_{n}=\sqrt{2P}/\sqrt{N}$ such that $\mathbf{S}=\sqrt{2P}/\sqrt{N}\mathbf{1}_{N}$. We will denote as UP this non-adaptive waveform strategy characterized by $\mathbf{S}=\sqrt{2P}/\sqrt{N}\mathbf{1}_{N}$ and $\mathbf{\Phi}=\mathbf{0}_{N}$. UP is suboptimal for $N>2$ and optimal for $N=2$ (as already found in Section \ref{Lagrangian_N2} when $A_0=A_1$), for which the inequality is replaced by an equality. Nevertheless for $N>2$, UP almost reaches the optimum obtained with Algorithm \ref{Algthm_OPT}, as confirmed in Section \ref{simulations}.

\par The value of $z_{DC}$ with the UP strategy, simply denoted as $z_{DC,UP}$, can be thought of as a lower bound on $z_{DC}(\mathbf{S}^{\star},\mathbf{\Phi}^{\star})$ (with optimal amplitude and phase strategy) in frequency-flat channels. Plugging $s_{n}=\sqrt{2P}/\sqrt{N}$ $\forall n$ into \eqref{z_DC_FF_general}, we get
\begin{equation}
z_{DC,UP}=k_2 A^2 R_{ant}  P+k_4 A^4 R_{ant}^2 \frac{2N^2+1}{2N}P^2 \label{UP_flat_N}
\end{equation} 
since that there are $N\left(2N^2+1\right)/3$ terms in the sum of \eqref{F}. 

\par In frequency-flat channels, $A\approx \sum_l \alpha_l e^{j\xi_l}$. Taking the expectation over $A$, $\bar{z}_{DC,UP}=\mathcal{E}\left\{z_{DC,UP}\right\}$ is written as
\begin{align}
\bar{z}_{DC,UP}&=k_2 R_{ant}  P+2 k_4 R_{ant}^2 \frac{2N^2+1}{2N}P^2\nonumber\\
&\stackrel{N\nearrow}{\approx}k_2 R_{ant} P+2 k_4 R_{ant}^2 N P^2\label{bar_z_DC_UP}
\end{align}
since $\mathcal{E}\left\{A^2\right\}=\sum_{l}\beta_l=1$ and $\mathcal{E}\left\{A^4\right\}=2 \sum_l \beta_l^2+2\sum_{l}\sum_{l'\neq l} \beta_l \beta_{l'}=2$ by making use of the moments of an exponential distribution ($\mathcal{E}\left\{\alpha_l^4\right\}=2\beta_l^2$).
\par Equations \eqref{UP_flat_N} and \eqref{bar_z_DC_UP} suggest that $z_{DC,UP}$ and $\bar{z}_{DC,UP}$ (and therefore $z_{DC}(\mathbf{S}^{\star},\mathbf{\Phi}^{\star})$) linearly increase with $N$ in frequency-flat channels. This is remarkable as it is achieved with a fixed waveform (non-adaptive to the CSI) and therefore without CSI feedback. We also note that the linear increase originates from the non-linearity of the rectifier as it only appears in the fourth order term. On the contrary, the transmission with a single sinewave ($N=1$) or with the ASS strategy would perform significantly worse with $z_{DC,SS/ASS}=k_2 A^2 R_{ant} P+\frac{3 k_4}{2} A^4 R_{ant}^2 P^2$ and $\bar{z}_{DC,SS/ASS}=k_2 R_{ant} P+3 k_4 R_{ant}^2 P^2$. The multisine waveform with uniform power allocation would achieve a relative gain over a single-sinewave strategy on a frequency-flat channel
that linearly increases with $N$. This gain highlights the potential of optimizing multisine waveforms and modeling the non-linearity of the rectifier.

\par Let us now look at multiple transmit antennas ($M\geq 1$). Since the channel is frequency flat, $\mathbf{h}_n=\mathbf{h}$, $\forall n$. Let us assume a simple strategy (denoted as UPMF) consisting in performing uniform power (UP) allocation in the frequency domain and matched beamforming (MF) in the spatial domain. We therefore write $\mathbf{w}_n=\sqrt{2P/N}\frac{\mathbf{h}^H}{\left\|\mathbf{h}\right\|}$, $\forall n$.
Making use of similar steps as in \eqref{UP_flat_N}, the harvested energy $z_{DC}$ writes as 
\begin{equation}\label{zdc_UP_MF_FF}
z_{DC,UPMF}=k_2 R_{ant} P\left\|\mathbf{h}\right\|^2+k_4 R_{ant}^2 \frac{2N^2+1}{2N}P^2 \left\|\mathbf{h}\right\|^4.
\end{equation} 
After averaging over the channel distribution and making use of the moments of a $\chi_{2M}^2$ random variable, we get
\begin{align}
\bar{z}_{DC,UPMF}&=k_2 R_{ant} P M+k_4 R_{ant}^2 P^2 \frac{2N^2+1}{2N} M\left(M+1\right)\nonumber\\
&\stackrel{N,M\nearrow}{\approx}k_2 R_{ant} P M+k_4 R_{ant}^2 P^2 N M^2.\label{UP_N_M}
\end{align}
The UPMF strategy enables an increase of $\bar{z}_{DC}$ proportionally to $N M^2$ and would rely on CSIT knowledge to perform spatial matched beamforming. While $M$ has an impact on both the second order and fourth order term, $N$ only appears in the fourth order term. Scaling law \eqref{UP_N_M} highlights that any increase of $\bar{z}_{DC,UPMF}$ by a factor 2 requires either increasing the number of sinewaves ($N$) by a factor 2 for a fixed number of transmit antennas ($M$) or increase the number of transmit antennas by a factor $\sqrt{2}$ for a fixed number of sinewaves. 

\par Let us now look at the presence of multiple rectennas ($U\geq 1$) and focus on $N\geq 1$ and $M=1$ for simplicity. Assuming the channels to each rectenna are identically distributed, the use of the UP strategy leads to an average harvested energy at rectenna $u$, $\bar{z}_{DC,UP,u}=\bar{z}_{DC,UP}$, that scales as \eqref{bar_z_DC_UP}.
Hence the sum energy $\bar{Z}_{DC,UP}=\sum_{u=1}^U \bar{z}_{DC,u}=U\bar{z}_{DC,UP}\stackrel{N\nearrow}{\approx} k_2 R_{ant} U P+2 k_4 R_{ant}^2 U N P^2$ linearly increases with $N$ and $U$. In frequency-flat channels with $U$ rectennas, the energy region $\mathcal{Z}_{DC}$ with UP strategy is a hypercube with each rectenna's harvested energy scaling linearly with $N$, i.e.\ the same quantity of energy as if it was alone in the system. Therefore adding more rectennas comes for free and does not compromise each rectenna's performance.   

\begin{remark} It may appear from \eqref{bar_z_DC_UP} and \eqref{UP_N_M} that taking $N$ to infinity would imply the harvested energy reaches infinity. The assumption behind the scaling law derivation is that the diode operates in the non-linear region, as discussed in Section \ref{rectifier_subsection}. If $N$ grows too large, the waveform peaks will ultimately have a very high amplitude and the diode will be forced into the linear region of operation, making the Taylor series expansion model and the scaling laws no longer applicable.
\end{remark}

\subsection{Frequency-Selective Channels}
We assume a frequency selective channel with $L>>1$ and frequencies $w_n$ far apart from each other such that the frequency domain circularly symmetric complex Gaussian random channel gains $h_{n,m}$ fade independently (phase and amplitude-wise) across frequencies and antennas.

\par Let us first consider $M=1$ and a waveform not adaptive to the CSI whose set of amplitude and phase is given by $\mathbf{S}$ and $\mathbf{\Phi}$. We write 
$z_{DC}(\mathbf{S},\mathbf{\Phi})=\frac{k_2}{2}R_{ant}\left[\sum_{n=0}^{N-1}X_n^2\right]+\frac{3k_4}{8}R_{ant}^2 \\ \left[\sum_{n=0}^{N-1}X_n^4+2\sum_{\mycom{n_0,n_1}{n_0\neq n_1}}X_{n_0}^2 X_{n_1}^2+ R \right]$
where $R$ contains all the remaining terms in the sum expansion \eqref{y_DC_4_1b}. Those terms are such that $\delta_{n_0}+\delta_{n_1}-\delta_{n_2}-\delta_{n_3}\neq 0$. We can compute the expectation of $z_{DC}(\mathbf{S},\mathbf{\Phi})$ over $\left\{h_n\right\}$. 
We note that $\mathcal{E}\left\{R\right\}=0$ because for any fixed phase of the waveform, quantities $\delta_{n_0}+\delta_{n_1}-\delta_{n_2}-\delta_{n_3}$ in $R$ would be uniformly distributed over $2\pi$ (since the phase of $h_n$ is uniformly distributed over $2\pi$) such that $\mathcal{E}\left\{\cos(\delta_{n_0}+\delta_{n_1}-\delta_{n_2}-\delta_{n_3})\right\}=0$.
Moreover, $\mathcal{E}\left\{X_n^2\right\}=s_{n}^2\mathcal{E}\left\{A_{n}^2\right\}=s_{n}^2$ and $\mathcal{E}\left\{X_n^4\right\}=s_{n}^4\mathcal{E}\left\{A_{n}^4\right\}=2 s_{n}^4$. Recalling the power constraint $\sum_{n=0}^{N-1}s_{n}^2=2P$, we can write
$\sum_{n=0}^{N-1}\mathcal{E}\left\{X_n^4\right\}+2\sum_{\mycom{n_0,n_1}{n_0\neq n_1}}\mathcal{E}\left\{X_{n_0}^2\right\}\mathcal{E}\left\{X_{n_1}^2\right\}=2\left[\sum_{n_0}s_{n_0}^2\right]\left[\sum_{n_1}s_{n_1}^2\right]=8P^2$,
therefore leading to
\begin{equation}\label{fixed_freq_fading}
\bar{z}_{DC}=\mathcal{E}\left\{z_{DC}\right\}=k_2 R_{ant}P+3k_4 R_{ant}^2 P^2.
\end{equation}
This highlights that in the presence of frequency-selective Rayleigh fading channels (with $L>>1$), $\bar{z}_{DC}$ is independent of $N$ and the waveform design, i.e.\ any fixed multisine waveform would achieve the same $\bar{z}_{DC}$. In the absence of CSIT, transmitting over a single sinewave ($N=1$) is enough in frequency-selective channels. In the presence of multiple rectennas, the sum energy writes as $\bar{Z}_{DC}=U\bar{z}_{DC}$.

\par Let us consider the same frequency-selective channel but now assume an adaptive waveform, namely the ASS strategy \eqref{solution_2nd_order} (still with $M=1$), allocating all transmit power to the sinewave corresponding to the strongest channel gain. We compute the expectation of $z_{DC}$ over $\left\{h_n\right\}$ as
\begin{equation}
\bar{z}_{DC,ASS}=\frac{k_2}{2}R_{ant}2P\mathcal{E}\left\{E_{max}\right\}+\frac{3k_4}{8}R_{ant}^2 4 P^2\mathcal{E}\left\{E_{max}^2\right\}.
\end{equation}
where $E_{max}=\max_n A_n^2$. Since $A_n^2\sim \textnormal{EXPO}(1)$, the pdf of $E_{max}$ simply is $f_{E_{max}}(x)=N e^{-x}\left(1-e^{-x}\right)^{N-1}$. Using \cite{Gradshteyn:2007}, $\mathcal{E}\left\{E_{max}\right\}=H_N$ and $\mathcal{E}\left\{E_{max}^2\right\}=2S_N$ with
\begin{align}
H_N&=N \left[\sum_{k=0}^{N-1}(-1)^{k+N-1}\small{\left(\begin{array}{c}N-1 \\ k\end{array}\right)}\frac{1}{\left(N-k\right)^2}\right],\\
S_N&=N \left[\sum_{k=0}^{N-1}(-1)^{k+N-1}\small{\left(\begin{array}{c}N-1 \\ k\end{array}\right)}\frac{1}{\left(N-k\right)^3}\right]
\end{align}
and we simply obtain
\begin{equation}
\bar{z}_{DC,ASS}=k_2R_{ant}P H_N+3k_4 R_{ant}^2 P^2 S_N.
\end{equation}
After some calculations, it can be shown that
\begin{equation} 
H_N=\sum_{k=1}^N \frac{1}{k}=\log N +\gamma+ \epsilon_N\stackrel{N\nearrow}{\approx} \log N + \gamma
\end{equation} 
with $\gamma$ the Euler-Mascheroni constant and $\epsilon_N$ scales as $\frac{1}{2N}$. 
Similarly, after some calculations, we can show that 
\begin{align}
S_N&=\sum_{k=1}^N \frac{1}{k} H_k=\sum_{k=1}^N\frac{\log k}{k}+ \gamma H_N+ \sum_{k=1}^N\frac{\epsilon_k}{k},\nonumber\\
&\stackrel{N\nearrow}{\approx}\frac{\log^2 N}{2}+\gamma_1+\gamma \log N+\gamma^2+ \sum_{k=1}^N\frac{\epsilon_k}{k},
\end{align}
where $\gamma_1$ is the Stieltjes constant. This shows that $H_N\approx \log N$ and $S_N\approx \frac{\log^2 N}{2}$.
We can now write
\begin{equation}
\bar{z}_{DC,ASS}\stackrel{N\nearrow}{\approx} k_2R_{ant}P \log N + \frac{3}{2}k_4 R_{ant}^2 P^2 \log^2 N.
\end{equation}
Thanks to the frequency selectivity, the ASS strategy enables an increase of the second order and fourth order terms proportionally to $\log N$ and $\log^2 N$, respectively.

\par Looking now at the UPMF strategy $\mathbf{w}_n=\sqrt{2P/N}\mathbf{h}_n^H/\left\|\mathbf{h}_n\right\|$ (for $N,M\geq 1$), we can write
\begin{equation}
\bar{z}_{DC,UPMF}=k_2 R_{ant} P M+\frac{3}{2}k_4 R_{ant}^2 \frac{P^2}{N^2} W.\label{bar_zdc_UP_MF_FS}
\end{equation}
where $W= \sum_{\mycom{n_0,n_1,n_2,n_3}{n_0+n_1=n_2+n_3}} \mathcal{E}\left\{\left\|\mathbf{h}_{n_0}\right\|\left\|\mathbf{h}_{n_1}\right\|\left\|\mathbf{h}_{n_2}\right\|\left\|\mathbf{h}_{n_3}\right\|\right\}$.
\par We can now lower and upper bound \eqref{bar_zdc_UP_MF_FS}. A lower bound is obtained by noting that 
$\mathcal{E}\big\{\prod_{j=0}^3\left\|\mathbf{h}_{n_j}\right\|\big\}\geq \prod_{j=0}^3\mathcal{E}\big\{\left\|\mathbf{h}_{n_j}\right\|\big\}$,
$\forall n_0,n_1,n_2,n_3$. Equality holds when $n_0\neq n_1 \neq n_2 \neq n_3$ due to the independence between channel gains in the frequency domain. Since $\left\|\mathbf{h}_{n}\right\|^2\sim \chi_{2M}^2$, we can compute $\mathcal{E}\left\{\left\|\mathbf{h}_{n}\right\|\right\}=\Gamma\left(M+\frac{1}{2}\right)/\Gamma\left(M\right)$. The lower bound is simply obtained from $W\geq \left(\Gamma\left(M+\frac{1}{2}\right)/\Gamma\left(M\right)\right)^4N \left(2N^2+1\right)/3 $. 
Noting $\mathcal{E}\left\{\left\|\mathbf{h}_{n_0}\right\|\left\|\mathbf{h}_{n_1}\right\|\left\|\mathbf{h}_{n_2}\right\|\left\|\mathbf{h}_{n_3}\right\|\right\}\leq \mathcal{E}\big\{\left\|\mathbf{h}_{n_0}\right\|^4\big\}=M\left(M+1\right)$, we also obtain the upper bound by writing $W\leq  M\left(M+1\right)N \left(2N^2+1\right)/3$.
Noting that $\lim_{M\rightarrow\infty}\frac{\Gamma\left(M+\alpha\right)}{\Gamma\left(M\right)M^{\alpha}}=1$ ($\alpha \in \mathbb{R}$), both lower and upper bounds have the same scaling law for $N,M$ growing large such that
\begin{equation}
\bar{z}_{DC,UPMF}\stackrel{N,M\nearrow}{\approx} k_2 R_{ant} P M+k_4 R_{ant}^2 P^2 N M^2.
\end{equation}
This is the same scaling law as \eqref{UP_N_M} in frequency flat channels. For $M=1$, if the fourth order term is dominant or if $N$ is large enough, the UPMF strategy\footnote{For $M=1$, UPMF should not be confused with UP. UPMF is an adaptive waveform that relies on CSIT knowledge to match the channel phases on each sinewave while UP is a non-adaptive waveform with null phases.} clearly outperforms the ASS strategy (i.e.\ linear versus log squared increase in $N$). On the other hand, if the second order term is dominant, the ASS strategy outperforms the UPMF strategy.

\par Table \ref{scaling_law_summary} summarizes the scaling laws for adaptive (based on CSIT) and non-adaptive (no CSIT) waveforms in frequency-flat and frequency-selective channels. We note again that for $M=1$ a linear increase with $N$ is achievable without CSIT in frequency-flat channels, while the same increase would require CSIT knowledge in frequency-selective channels. We also note that a linear model-based design leads to significantly lower scaling laws than the non-linear model-based design for frequency-flat and frequency-selective channels. This really highlights the importance of modeling higher order terms in the Taylor expansion. 

\begin{table*}
\caption{Summary of Scaling Laws.}
\centering
\begin{tabular}{|p{1.5cm}|p{3cm}||p{4cm}|p{5cm}|}
\hline \textbf{Waveform} & $N,M$	& \textbf{Frequency-Flat (FF)} &	\textbf{Frequency-Selective (FS)} \\
\hline
\hline \textbf{No CSIT}	&  &  &  \\
\hline $\bar{z}_{DC,SS}$	& $N=1,M=1$ & $k_2 R_{ant} P+3 k_4 R_{ant}^2 P^2$ &  \\
\hline $\bar{z}_{DC,UP}$	& $N >> 1,M=1$ & $k_2 R_{ant} P+2 k_4 R_{ant}^2 N P^2$	& $k_2 R_{ant} P+3 k_4 R_{ant}^2 P^2$ \\
\hline $\bar{Z}_{DC,UP}$	& $N\geq 1,M=1, U\geq 1$ & $U \bar{z}_{DC,UP}$ ($v_u=1$, $\forall u$)	& $U \bar{z}_{DC,UP}$ ($v_u=1$, $\forall u$) \\
\hline \textbf{CSIT} &  &  & \\
\hline $\bar{z}_{DC,ASS}$	& $N >> 1,M=1$ & $k_2 R_{ant} P+3 k_4 R_{ant}^2 P^2$	& $k_2R_{ant}P \log N + \frac{3}{2}k_4 R_{ant}^2 P^2 \log^2 N$ \\
\hline $\bar{z}_{DC,UPMF}$	& $N >> 1,M = 1$ & $k_2 R_{ant} P+2k_4 R_{ant}^2 P^2 N $	& $\geq k_2 R_{ant} P +\pi^2/16 k_4 R_{ant}^2 P^2 N, \hspace{0.55cm}\leq k_2 R_{ant} P +2 k_4 R_{ant}^2 P^2 N$ \\
\hline $\bar{z}_{DC,UPMF}$	& $N >> 1,M >> 1$ & $k_2 R_{ant} P M+k_4 R_{ant}^2 P^2 N M^2$	& $k_2 R_{ant} P M+k_4 R_{ant}^2 P^2 N M^2$ \\
\hline
\end{tabular}
\label{scaling_law_summary}
\end{table*}

\subsection{Large-Scale Multi-Sine Multi-Antenna WPT} 
The previous scaling laws highlight the benefits of a large-scale multisine multi-antenna architecture. This is reminiscent of Massive MIMO in communication. The large dimension enables to significantly simplify the waveform design. The spatial matched beamformer \eqref{opt_weight}, $\mathbf{w}_n=s_n \mathbf{h}_n^H/\left\|\mathbf{h}_n\right\|$ (with $\sum_{n=0}^{N-1}s_n^2=2P$), would induce channel hardening on sinewave $n$ such that by the law of large number $\lim_{M\rightarrow \infty} \left\|\mathbf{h}_{n}\right\|/\sqrt{M}=1$ and 
\begin{equation}
z_{DC}\stackrel{M\nearrow}{\approx}k_2 R_{ant} P M +\frac{3}{8}k_4 R_{ant}^2 M^2 F
\end{equation}
where $F$ is defined in \eqref{F}. $z_{DC}$ can now be maximized by using the optimal power allocation for frequency-flat channels. The suboptimal UP would be a good alternative. This leads to a very low complexity waveform design for large-scale WPT.
 
\section{Performance Evaluations}\label{simulations}

\par We consider two types of performance evaluations, the first one is based on the simplified non-linear model introduced in Section \ref{section_EH}, while the second one relies on an actual and accurate modelling of the rectenna in PSpice. 

\subsection{Non-Linear Model-Based Performance Evaluations}
\par The first type of evaluations consists in displaying $z_{DC}$ averaged over many channel realizations for various waveforms. To that end, we assume a fourth order Taylor expansion and therefore consider the following metric $z_{DC}(\mathbf{S},\mathbf{\Phi})=k_2 R_{ant} \mathcal{E}\left\{y(t)^2\right\}+k_4 R_{ant}^2 \mathcal{E}\left\{y(t)^4\right\}$ with $k_2=0.0034$, $k_4=0.3829$ and $R_{ant}=50\Omega$.  

\par
\begin{figure}
\centerline{\includegraphics[width=0.9\columnwidth]{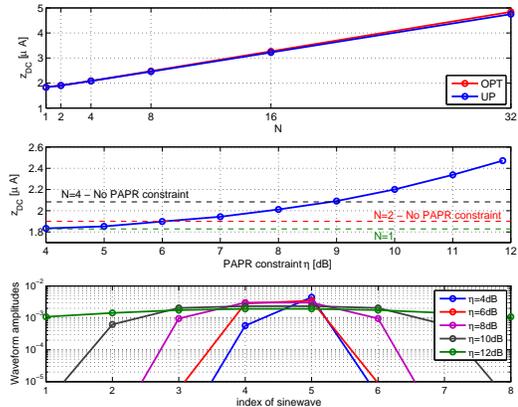}}
  \caption{$z_{DC}$ as a function of $N$ (top) and PAPR constraint $\eta$ for $N=8$ (middle). WPT waveform amplitude as a function of $\eta$ for $N=8$ (bottom). No wireless channel is assumed, i.e.\ $A=1$ and $\bar{\psi}=0$, and $M=1$.}
  \label{z_DC_nochannel_OPT_UP}
\end{figure}

 We first consider a single rectenna scenario where the wireless channel is omitted, i.e.\ $A=1$ and $\bar{\psi}=0$ (representing a frequency flat channel) and a single transmit antenna. The received power, i.e.\ input power to the rectenna, is fixed to -20dBm. Fig \ref{z_DC_nochannel_OPT_UP} (top) confirms that in a frequency flat channel, $z_{DC}$ with UP is close to that achieved by OPT, obtained from Algorithm \ref{Algthm_OPT}. Fig \ref{z_DC_nochannel_OPT_UP} (middle) investigates the impact of PAPR constraint on $z_{DC}$ with the optimized waveform for $N=8$ using Algorithm \ref{Algthm_OPT_PAPR}. Fig \ref{z_DC_nochannel_OPT_UP} (bottom) illustrates the corresponding shape of the waveform amplitudes $s_n$ across frequencies for various PAPR constraints $\eta$. As $\eta$ decreases, the allocation of power decreases on the side frequencies and concentrates more on the center frequencies. For large $\eta$, the optimized waveform never exactly reaches the UP waveform. Center frequencies get slightly larger magnitudes, which explains a slight increase in $z_{DC}$ of OPT over UP in Fig \ref{z_DC_nochannel_OPT_UP} (top).

\par We now evaluate the performance of WPT waveforms in a single rectenna scenario representative of a WiFi-like environment at a center frequency of 5.18GHz with a 36dBm transmit power, isotropic transmit antennas (i.e.\ EIRP of 36dBm for $M=1$), 2dBi receive antenna gain and 58dB path loss in a large open space environment with a NLOS channel power delay profile with 18 taps obtained from model B \cite{Medbo:1998b}. Taps are modeled as i.i.d.\ circularly symmetric complex Gaussian random variables, each with an average power $\beta_l$. The multipath response is normalized such that $\sum_{l=1}^{18} \beta_l=1$. With one transmit antenna, this leads to an average received power of -20dBm ($10\mu W$). Equivalently, this system model can be viewed as a transmission over the aforementioned normalized multipath channel with an average transmit power fixed to -20dBm. The frequency gap is fixed as $\Delta_w=2\pi\Delta_f$ with $\Delta_f=B/N$ and $B=1,10$MHz. The $N$ sinewaves are centered around 5.18GHz.

\par
\begin{figure}
\centerline{\includegraphics[width=0.9\columnwidth]{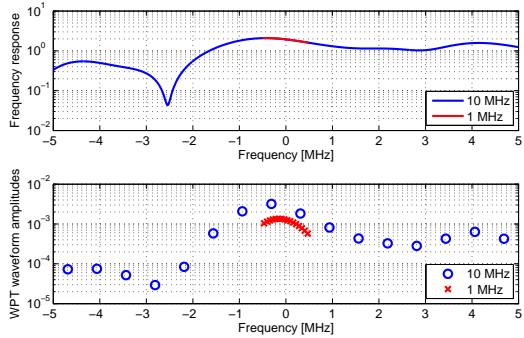}}
  \caption{Frequency response of the wireless channel and WPT waveform magnitudes ($N=16$) for 1 MHz and 10 MHz bandwidths.}
  \label{Freq_response_channel1}
\end{figure}

 In Fig \ref{Freq_response_channel1}, we illustrate the effect of frequency selectivity on the shape of the transmit waveform obtained using Algorithm \ref{Algthm_OPT}. Fig \ref{Freq_response_channel1} (top) illustrates the frequency response of one realization of the multipath channel over 1MHz and 10 MHz bandwidth. Fig \ref{Freq_response_channel1} (bottom) displays the magnitude of the waveform optimized for $N=16$ (Algorithm \ref{Algthm_OPT}) over such a channel realization. Interestingly, the optimized waveform has a tendency to allocate more power to frequencies exhibiting larger channel gains. This is reminiscent of the water-filling power allocation strategy in communication. This observation also suggests a suboptimal low complexity waveform design that would allocate power proportionally to the channel strength. For comparison, recall that the ASS waveform, motivated by the linear model, would allocate all power to the frequency corresponding to the strongest channel gain.

\par We now evaluate the performance gain of the adaptive optimized (OPT) waveform (Algorithm \ref{Algthm_OPT}) versus three baselines: a non-adaptive waveform not relying on CSIT and two adaptive waveforms relying on CSIT but not requiring the optimization of Section \ref{section_waveform}. From the scaling law analysis, a suitable choice of non-adaptive waveform for single antenna WPT is UP. We therefore choose the non-adaptive baseline waveform as $\phi_{n,m}=0$ and $s_{n,m}=\sqrt{2P}/\sqrt{NM}$ $\forall n,m$. Motivated by the observations made in Fig \ref{Freq_response_channel1}, the first adaptive baseline waveform is chosen as a matched filter (MF) allocating power to all sinewaves but proportionally to the channel strength, i.e.\ $\phi_{n,m}=-\bar{\psi}_{n,m}$ and $s_{n,m}=c A_{n,m}$ with $c$ a constant to guarantee the power constraint. Hence the difference between the optimized waveform and the one based on MF lies in a different choice of amplitudes. The second adaptive baseline waveform is the ASS, designed according to the linear model. 

\par
\begin{figure}
\centerline{\includegraphics[width=0.9\columnwidth]{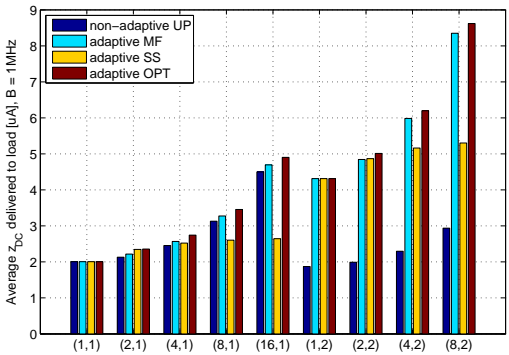}}
  \caption{Average $z_{DC}$ as a function of $(N,M)$ with $B=1$MHz.}
  \label{z_DC_results_1MHz}
\end{figure}

\begin{figure}
\centerline{\includegraphics[width=0.9\columnwidth]{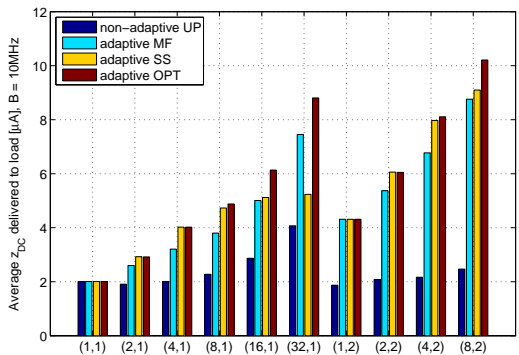}}
  \caption{Average $z_{DC}$ as a function of $(N,M)$ with $B=10$MHz.}
  \label{z_DC_results_10MHz}
\end{figure}

 Fig \ref{z_DC_results_1MHz} and \ref{z_DC_results_10MHz} display $z_{DC}$ averaged over many channel realizations as a function of $(N,M)$ for two bandwidths $B=1 MHz$ and $B=10 MHz$, respectively. We make the following observations. First, for small bandwidth ($B=1$MHz), the UP non-adaptive waveform performs pretty well in the presence of a single transmit antenna ($M=1$), confirming that for channels with little frequency selectivity, CSI feedback is not needed. On the other hand, for larger bandwidth ($B=10$MHz), the non-adaptive waveform is clearly outperformed by the adaptive waveforms, therefore highlighting the usefulness of CSI feedback in WPT even with a single transmit antenna. Second, for small bandwidth, the ASS waveform is significantly outperformed by the UP waveform for $M=1$, despite the fact it requires CSI knowledge at the Transmitter. For larger bandwidth, the ASS waveform benefits from the channel frequency selectivity to get close performance to OPT for small $N$. As $N$ increases, the ASS waveform is however clearly outperformed by the adaptive MF and OPT waveforms. This highlights the inaccuracy of the linear model in characterizing the rectifier and the inefficiency of the linear model-based design. The inefficiency is particularly severe as $N$ increases irrespectively of the bandwidth. These observations confirm the predictions made from the scaling laws in Table \ref{scaling_law_summary}. Third, OPT outperforms all waveforms in all configurations. Fourth, MF is a good alternative to OPT, at least with small bandwidth, and does not require any optimization. For larger bandwidth, OPT shows a non-negligible gain over MF as $N$ increases.



\par
\begin{figure}
\centerline{\includegraphics[width=0.9\columnwidth]{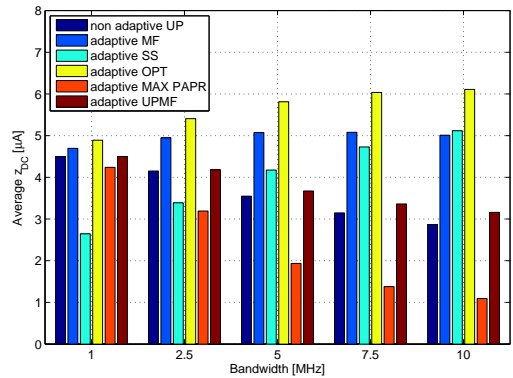}}
  \caption{Effect of Bandwidth $B$ on $z_{DC}$ for $N=16$ and $M=1$.}
  \label{Average_z_DC_N16_M1}
\end{figure}

 Fig \ref{Average_z_DC_N16_M1} further analyzes the sensitivity of $z_{DC}$ to the bandwidth for a fixed number of sinewaves $N=16$ and various waveforms. Waveforms relying on uniform power allocation such as non-adaptive UP and adaptive UPMF experience some loss as the bandwidth increases and the channel becomes more frequency selective. On the other hand, adaptive OPT and adaptive SS benefit from the frequency selectivity by favouring the strongest sinewave(s). In \cite{Collado:2014}, experiments show that waveforms with high peak to average power ratio (PAPR) increase RF-to-DC conversion efficiency. The conclusion was drawn for various waveforms (OFDM, white noise, chaotic) that were not designed or optimized for WPT. Following this observation, we investigate whether designing waveforms so as to maximize the PAPR at the input of the rectenna, after the wireless channel, is a suitable approach. The adaptive waveform MAX PAPR in Fig \ref{Average_z_DC_N16_M1} is designed following this philosophy. It uses the same phases as OPT but inverts the channel such that at the input to the rectifier, the waveform appears as an in-phase multisine with uniform power allocation (which is known to have the maximum PAPR of $10 \log_{10} \left(2N\right)$ dB). This is mathematically formulated by choosing $s_n^2=C/A_n^2$ where $C$ is a constant to satisfy the transmit power constraint. Results show that this is a rather inefficient waveform design strategy. This originates from the relatively low magnitude of the waveform peaks due to the excessive amount of power wasted in inverting the wireless power to guarantee the maximum PAPR at the input of the rectenna. Note also that non-adaptive UP would lead to the highest transmit PAPR (i.e. PAPR of the transmit waveform, before the wireless channel) due the uniform allocation across 16 in-phase sinewaves. OPT on the other hand has a transmit PAPR always lower than UP despite providing higher $z_{DC}$.

\par
\begin{figure}
\centerline{\includegraphics[width=0.9\columnwidth]{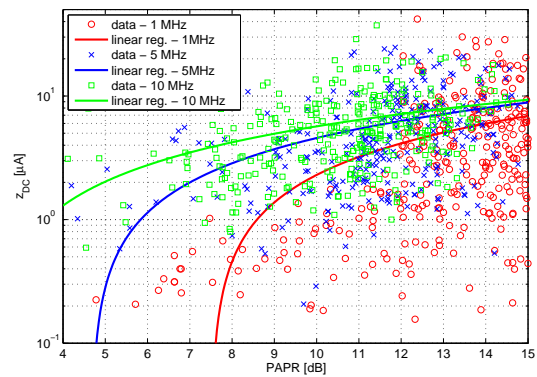}}
  \caption{$z_{DC}$ versus transmit PAPR for $N=16$ and $M=1$.}
  \label{z_DCvsPAPR_N16}
\end{figure}

 Fig \ref{z_DCvsPAPR_N16} further investigates the impact of PAPR on the performance of the optimized multisine waveforms. It considers the OPT waveform with 16 sinewaves uniformly spread over 3 different bandwidths. $z_{DC}$ is plotted against the PAPR of the transmit waveform for each realization of the multipath channel, along with some linear regression fit. It is noted that there is some positive correlation between $z_{DC}$ and PAPR, especially for small bandwidths. As the bandwidth increases and the wireless channel becomes more frequency selective, the optimized waveform has a tendency to allocate less power to the weakest channels, therefore leading to lower PAPR. This explains why as the bandwidth increases, the correlation between DC current and PAPR reduces. 

\begin{figure}
\centerline{\includegraphics[width=0.9\columnwidth]{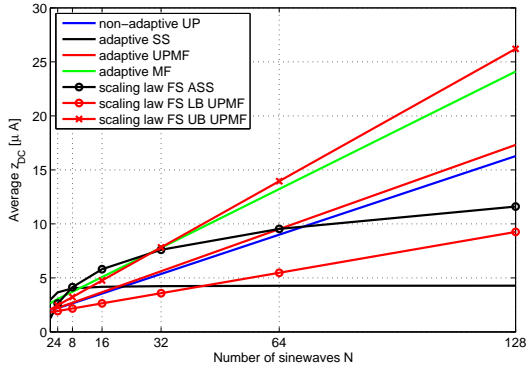}}
  \caption{Effect of $N$ on $z_{DC}$ for $M=1$ and 5 MHz bandwidth.}
  \label{average_z_DC_5MHz_scalinglaws}
\end{figure}

\par Fig \ref{average_z_DC_5MHz_scalinglaws} reveals the performance of a large-scale multisine WPT using 4 suboptimal (though low complexity) waveforms (UP, ASS, UPMF and MF\footnote{The OPT waveform is not computed given the high computational complexity of the optimization for large $N$. This calls for future research on alternative optimization methods for large-scale waveforms.}) for $M=1$ and 5 MHz bandwidth. The linear model-based ASS is significantly outperformed by the non-linear model-based design as $N$ grows large. The scaling laws for ASS and UPMF over frequency-selective (FS) channels in Table \ref{scaling_law_summary} are also displayed for comparison.

\subsection{Accurate and Realistic Performance Evaluations}
\par
\begin{figure}
\centerline{\includegraphics[width=0.8\columnwidth]{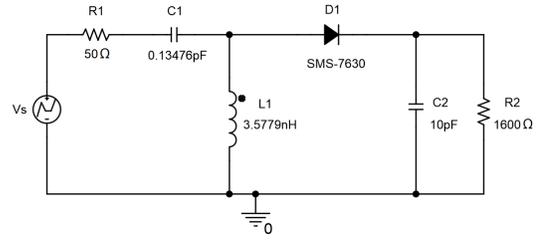}}
  \caption{Rectenna with a single diode and a L-matching network used for PSpice evaluations with $B=10$MHz.}
  \label{circuit}
\end{figure}

 The second type of evaluations is based on an accurate modeling of the rectenna in PSpice in order to validate the waveform optimization and the rectenna non-linearity model. To that end, the waveforms after the wireless channel have been used as inputs to the realistic rectenna of Fig \ref{circuit} designed for an input power of -20dBm. The circuit contains an L-matching network \cite{Pinuela:2013} to guarantee a good matching between the rectifier and the antenna and to minimize the impedance mismatch due to variations in frequency and input power level of the input signal. The values of the capacitor $\textnormal{C1}$ and the inductor $\textnormal{L1}$ are optimized to match the antenna impedance to the average input impedance of the rectifier resulting from an input signal composed of 4 sinewaves and spread across $B=10$MHz. Using ADS Harmonic Balance simulation, the rectifier impedance is measured at the 4 sinewave frequencies during a few iterations, and conjugate matching is performed to match the antenna to the average rectifier impedance at each iteration until the impedance mismatch error is minimised. $\textnormal{Vs}=v_s(t)=2 y(t) \sqrt{R_{ant}}$ is set as the voltage source. Taking the optimized waveform as an example, for a given channel realization, Algorithm \ref{Algthm_OPT} is used to derive the optimal waveform weights, which are then used to generate in Matlab the waveform $y(t)$ as in \eqref{y_t} (after the wireless channel). Several periods of the signal are generated such that $t=0,\ldots,c \Delta_t$, with $c$ a positive integer chosen sufficiently large to make sure that the rectifier is in the steady-state response mode and $\Delta_t=1/\Delta_f$ the period of the waveform. Quantity $y(t)$ is stored and then fed into the PSpice circuit simulator to generate the voltage source $\textnormal{Vs}$ in Fig \ref{circuit}. The antenna and load impedances are set as $\textnormal{R1}=R_{ant}=50\Omega$ and $\textnormal{R2}=R_L=1600\Omega$, respectively. The output capacitor is chosen as $\textnormal{C2}=C_{out}=100$pF for $B=1$MHz and $\textnormal{C2}=C_{out}=10$pF for $B=10$MHz so that the output DC power is maximized and the rate of charge and discharge of $C_{out}$ is maintained in proportion to the period of the waveform, i.e.\ for evaluations with $B=1$MHz, $\textnormal{C2}$ is replaced by a 100pF capacitor in Fig \ref{circuit}. 

\par
\begin{figure}
\centerline{\includegraphics[width=0.9\columnwidth]{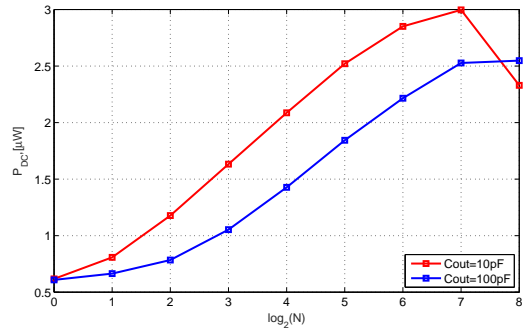}}
  \caption{$P_{DC}$ as a function of $N$ for $B=10$MHz. No wireless channel is assumed, i.e.\ $A=1$ and $\bar{\psi}=0$, and $M=1$.}
  \label{Pdc_N_Cout_10p_100p}
\end{figure}

 Fig \ref{Pdc_N_Cout_10p_100p} illustrates the increase of the harvested DC power as a function of $N$ for a single transmit antenna and assuming no wireless channel, i.e.\ $A=1$ and $\bar{\psi}=0$ (representing a frequency flat channel). The harvested DC power is not a monotonically increasing function contrary to what was observed in Fig \ref{z_DC_nochannel_OPT_UP} with $z_{DC}$. This is explained by the fact that the rectenna has been optimized for 4 sinewaves. For $B=10$MHz and $N=4$, $C_{out}=10$pF was found appropriate. Nevertheless, as $N$ increases, for a fixed $B$, $\Delta_f$ decreases, which affects the rate of charge and discharge of the output capacitor. This shows that $C_{out}$ (but also the load and the matching network) should ideally be adjusted as a function of $N$. We indeed notice that for large $N$, a larger capacitor of 100pF is better than 10pF. It is worth noting even if the rectenna design changes as a function of $N$, beyond a certain $N$, the peak of the voltage at the input of the diode would be higher than the diode breakdown voltage (2V for SMS7630), which would cause a sharp decrease in efficiency.

\par
\begin{figure}
\centerline{\includegraphics[width=0.9\columnwidth]{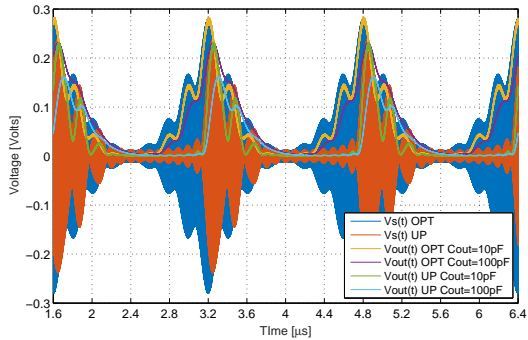}}
    \caption{$v_{s}(t)$ and $v_{out}(s)$ over the wireless channel of Fig \ref{Freq_response_channel1} with OPT and UP waveforms for $N=16$, $B=10$ MHz and $C_{out}=10$pF,100pF.}
\label{Waveform_shape}
  \label{vin}
\end{figure}

 In Fig \ref{Waveform_shape}, considering the channel impulse response of Fig \ref{Freq_response_channel1}, we illustrate the time-domain evolution of the input and output voltages (in the form of $v_s(t)$ and $v_{out}(t)$) for the OPT and UP waveforms (with $N=16$ and $B=10$MHz). We also illustrate the effect of the output capacitance $C_{out}$ on the performance. Large peaks in the input voltage occur with a periodicity $\Delta_t=1/\Delta_f=N/B=1.6\mu$s. Output voltage is not flat contrary to what is expected with an ideal rectifier (as used in the non-linear model of Section \ref{section_EH}). This is due to the finite $R_{L}C_{out}$ chosen in the simulated (and optimized) rectifier of Fig \ref{circuit}. We note that a larger $C_{out}$ leads to a smoother output voltage and a better discharging behaviour but a slower charging time and lower output peak voltages. A good value for $C_{out}$ results from a compromise between those conflicting mechanisms that explains why a finite $C_{out}$ is needed in practice. We also note that the OPT waveform leads to a higher output voltage than that obtained with the UP waveform. The harvested DC output power with $C_{out}=100$pF is given by $2.3281\mu$W and $6.4157\mu$W, for UP and OPT, respectively. With $C_{out}=10$pF, the harvested DC output power is slightly higher and given by $2.9435\mu$W and $6.9387\mu$W, respectively.

\par
\begin{figure}
\centerline{\includegraphics[width=0.9\columnwidth]{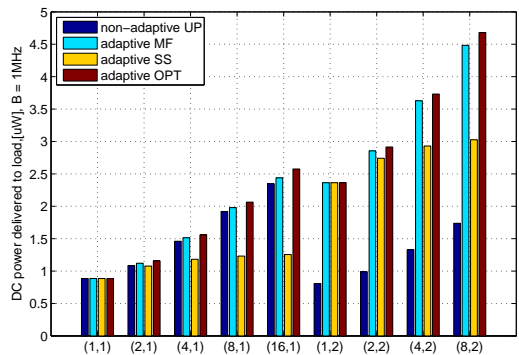}}
  \caption{Average DC power as a function of $(N,M)$ with $B=1$MHz.}
  \label{pspice_results_1MHz}
\end{figure}

\begin{figure}
\centerline{\includegraphics[width=0.9\columnwidth]{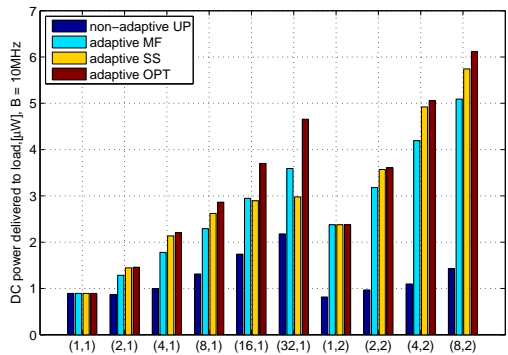}}
  \caption{Average DC power as a function of $(N,M)$ with $B=10$MHz.}
  \label{pspice_results_10MHz}
\end{figure}

 Fig \ref{pspice_results_1MHz} and \ref{pspice_results_10MHz} display the average harvested DC output power for $B=1 $MHz and $B=10$MHz (using the same channel realizations as those used in Fig \ref{z_DC_results_1MHz} and \ref{z_DC_results_10MHz}), respectively. We make important observations. First, the results confirm the observations made in Fig \ref{z_DC_results_1MHz} and \ref{z_DC_results_10MHz} and validate the rectenna non-linearity model\footnote{This does not mean that the model $z_{DC}$ is accurate enough to predict the rectifier output DC power using $R_L \left(k_0+z_{DC}\right)^2$.} and the waveform optimization. There is indeed a good match between the behavior predicted from the analytical nonlinear model and the one observed from the PSpice simulations. Second, they highlight the significant (and increasing as $N,M$ grow) gains achieved by the nonlinear model-based design over the linear model-based design. Third, they highlight that the linear model does not characterize correctly the rectenna behavior, which leads to an inefficient multisine waveform design. Indeed, if the linear model had accurately characterized the rectifier behavior, the ASS waveform would have provided the highest average DC power over all other waveforms. It is clearly not the case. The behaviour observed from Fig \ref{pspice_results_1MHz} and \ref{pspice_results_10MHz} cannot be explained based on the linear model. In Fig \ref{pspice_results_1MHz}, with $M=1$, ASS (requiring CSIT) is even outperformed by non-adaptive UP (not requiring CSIT). 
\par Results in Fig \ref{pspice_results_1MHz} and \ref{pspice_results_10MHz} can also be viewed in terms of RF-to-DC conversion efficiency by dividing the harvested DC power by the average input power ($10\mu W$). For 10MHz and $M=1$, the RF-to-DC conversion efficiency of the OPT waveform is 9\%, 15\%, 22\%, 28\%, 37\% and 46\% for $N=1,2,4,8,16$ and 32, respectively.
\par It is worth noting in Fig \ref{pspice_results_1MHz} and \ref{pspice_results_10MHz} the effect of bandwidth on average DC power. The average DC power with a 10 MHz bandwidth is larger than that with a 1 MHz bandwidth. This comes from the increased channel frequency selectivity and the diode being turned on more often as $\Delta_f$ increases. When $N=1$, all waveforms obviously achieve the same performance.


\begin{remark} 
$n_o=4$ was used throughout the waveform design and $z_{DC}$ evaluations. More investigations are needed to assess the usefulness of even higher order terms ($n_o\geq 6$). In \cite{Boaventura:2011}, it was claimed that $n_o=4$ is the minimum order required to characterize the nonlinear mechanisms of the diode.
\end{remark}

\section{Conclusions and Future Works}\label{conclusions}
The paper looks at a WPT link optimization and derives a methodology to design and optimize multisine waveforms for WPT. Assuming the CSI is available to the transmitter, the waveforms result from a non-convex posynomial maximization problem and are shown through realistic simulations to provide significantly higher harvested DC power over various baseline waveforms under a fixed transmit power constraint. The results show the importance of accounting for the non-linearity of the rectifier in any design involving wireless power.

\par Due to the space limitation, there are many important and exciting research avenues unaddressed in this paper and left for future work. Some of them are highlighted below.
\par The waveform design problem addressed in this paper assumes $N$ sinewaves with a uniform frequency spacing $\Delta_f=B/N$ for a given spectrum bandwidth $B$. A fundamental question arising from this work is, given a spectrum bandwidth $B$, what is the best way to transmit power so as to maximize the output DC power? This would help understanding how to make the best use of the RF spectrum for WPT. This problem has been investigated for several decades in wireless communication but is an uncharted area in WPT. 
\par The work highlights the usefulness of adaptive waveforms and CSIT. The fundamentals of CSI acquisition/feedback in WPT remain largely unknown. Some interesting ideas along this line have appeared in \cite{Xu:2014}. However, the work relied on the linear model. It is unclear yet whether a similar approach can be used over the non-linear wireless power channel. 
\par The scaling laws and evaluations highlight the potentials of a promising architecture relying on large-scale multisine multi-antenna waveforms dedicated to WPT. This architecture would be to wireless power what massive MIMO is to communication. More results along this line can be found in \cite{Huang:2016}.
\par The work also highlights the importance of understanding and modeling the wireless power channel and formulating a complete link optimization (transmitter to rectenna) in order to design an efficient WPT architecture. Since WPT is the fundamental building block of various types of wireless powered systems (e.g.\ WPT, SWIPT, WPCN, backscatter communication), this motivates a bottom-up approach where any wireless powered system is based on a sound science-driven design of the underlying WPT. The waveform design and the rectifier non-linearity tackled in this paper therefore have direct consequences on the design of SWIPT, WPCN and backscatter communication. For instance, some preliminary results on SWIPT waveforms have been reported in \cite{Clerckx:2016}, where it is shown that the superposition of multisine and OFDM waveforms enlarges the rate-energy region compared to an OFDM-only transmission. This originates from the non-linearity of the rectifier and the fact that the OFDM waveform, due to the randomness of the information, is less efficient than a (deterministic) multisine waveform to convert RF power to DC power. 

\ifCLASSOPTIONcaptionsoff
  \newpage
\fi

\end{document}